\documentclass[a4paper,12pt]{article}
\usepackage{amsfonts,amsthm,amsmath,amssymb,graphicx}
\usepackage{braket}
\usepackage[utf8]{inputenc}
\usepackage{color}   %May be necessary if you want to color links
\usepackage{hyperref}
\usepackage{cite}

\hypersetup{
    colorlinks=true, %set true if you want colored links
    linktoc=all,     %set to all if you want both sections and subsections linked
    linkcolor=blue,  %choose some color if you want links to stand out
}
\numberwithin{equation}{section}
\usepackage{color}
\begin{document}

\newtheorem{definition}{Definition}[section]
\newcommand{\be}{\begin{equation}}
\newcommand{\ee}{\end{equation}}
\newcommand{\bea}{\begin{eqnarray}}
\newcommand{\eea}{\end{eqnarray}}
\newcommand{\LE}{\left[}
\newcommand{\R}{\right]}
\newcommand{\nn}{\nonumber}
\newcommand{\Tr}{\text{Tr}}
\newcommand{\N}{\mathcal{N}}
\newcommand{\G}{\Gamma}
\newcommand{\vf}{\varphi}
\newcommand{\LL}{\mathcal{L}}
\newcommand{\Op}{\mathcal{O}}
\newcommand{\HH}{\mathcal{H}}
\newcommand{\arctanh}{\text{arctanh}}
\newcommand{\up}{\uparrow}
\newcommand{\down}{\downarrow}
\newcommand{\ketbra}[1]{\left|#1\right>\left<#1\right|}
\newcommand{\rd}{\partial}
\newcommand{\de}{\partial}
\newcommand{\ba}{\begin{eqnarray}}
\newcommand{\ea}{\end{eqnarray}}
\newcommand{\db}{\bar{\partial}}
\newcommand{\we}{\wedge}
\newcommand{\ca}{\mathcal}
\newcommand{\lr}{\leftrightarrow}
\newcommand{\f}{\frac}
\newcommand{\s}{\sqrt}
\newcommand{\vp}{\varphi}
\newcommand{\hvp}{\hat{\varphi}}
\newcommand{\tvp}{\tilde{\varphi}}
\newcommand{\tp}{\tilde{\phi}}
\newcommand{\ti}{\tilde}
\newcommand{\ap}{\alpha}
\newcommand{\pr}{\propto}
\newcommand{\mb}{\mathbf}
\newcommand{\ddd}{\cdot\cdot\cdot}
\newcommand{\no}{\nonumber \\}
\newcommand{\la}{\langle}
\newcommand{\lb}{\rangle}
\newcommand{\ep}{\epsilon}
 \def\we{\wedge}
 \def\lr{\leftrightarrow}
 \def\f {\frac}
 \def\ti{\tilde}
 \def\ap{\alpha}
 \def\pr{\propto}
 \def\mb{\mathbf}
 \def\ddd{\cdot\cdot\cdot}
 \def\no{\nonumber \\}
 \def\la{\langle}
 \def\lb{\rangle}
 \def\ep{\epsilon}
\newcommand{\mcl}{\mathcal}
 \def\g{\gamma}
\def\tr{\text{tr}}

\begin{titlepage}
\thispagestyle{empty}

\begin{flushright}
%NORDITA-2015-137\\
%EFI-15-38\\
%YITP-15-118\\
\end{flushright}
\bigskip

\begin{center}
\noindent{\large \textbf{Geometry and Complexity of Path Integrals\\ in Inhomogeneous CFTs}}\\
 \vspace{2cm}
% \today \\
% \vspace{1cm}
 Pawe{\l} Caputa $^{a}$ and 
 \vspace{1cm}
 Ian MacCormack $^{b}$

% \vspace{1cm}
{\it $^{a}$ Faculty of Physics, University of Warsaw, ul. Pasteura 5, 02-093 Warsaw, Poland}

 {\it $^{b}$Kadanoff Center for Theoretical Physics, University of Chicago,\\
 Chicago, Illinois 60637, USA \\}

\vskip 4em
\end{center}
%%%%%%%%%%%%%%%%%%%%%%%%%%%%%
\begin{abstract}
In this work we develop the path integral optimization in a class of inhomogeneous 2d CFTs constructed by putting an ordinary CFT on a space with a position dependent metric. After setting up and solving the general optimization problem, we study specific examples, including the M\"obius, SSD and Rainbow deformed CFTs, and analyze path integral geometries and complexity for universal classes of states in these models. We find that metrics for optimal path integrals coincide with particular slices of $AdS_3$ geometries, on which Einstein's equations are equivalent to the condition for minimal path integral complexity. We also find that while leading divergences of path integral complexity remain unchanged, constant contributions are modified in a universal, position dependent manner. Moreover, we analyze entanglement entropies in inhomogeneous CFTs and show that they satisfy Hill's equations, which can be used to extract the energy density consistent with the first law of entanglement. Our findings not only support comparisons between slices of bulk spacetimes and circuits of path integrations, but also demonstrate that path integral geometries and complexity serve as a powerful tool for understanding the interesting physics of inhomogeneous systems.

\end{abstract}
%%%%%%%%%%%%%%%%%%%%%%%%
\end{titlepage} 
%%%%%%%%%%%%%%%%%%%%%%%%
\tableofcontents

%%%%%%%%%%%%%%%%%%%%%%%%
%%%%%%%%%%%%%%%%%%%%%%%%
\section{Introduction}
%%%%%%%%%%%%%%%%%%%%%%%%
%%%%%%%%%%%%%%%%%%%%%%%%
%(To be written at the end) Include:\\
%1. Geometry from quantum states, AdS/CFT, path integrals, Optimization, Continuous tensor %networks \cite{Milsted:2018yur}. Holographic Slices.\\
%2. Inhomogeneous CFTs, Their Geometry.\\
%3. Path integral complexity. How difficult it is to prepare states in inhom. CFTs form the PI %perspective?\\

It has been known for some time that quantum entanglement can be geometrized using tools from quantum information theory (see e.g. \cite{Bengtsson}). This paradigm proved to be very fruitful in the context of the AdS/CFT correspondence, where the Ryu-Takayanagi formula \cite{Ryu:2006bv} provides a direct link between entanglement in the boundary CFT and the areas of surfaces in the bulk Anti-de Sitter geometry. Still, it is likely that entanglement entropy is not sufficient to fully understand the bulk geometry \cite{Susskind:2014moa}. Moreover, we are not sure how much quantum information technology should be incorporated into CFTs in order to precisely extract the full holographic geometry and understand the mechanism behind AdS/CFT.

A promising hint in this puzzle has emerged from tensor network methods frequently used to represent quantum states in many body physics \cite{2008PhRvL.101k0501V,2012PhRvD..86f5007S}. By imposing certain entanglement properties on tensor networks, they can be used as toy models for holographic systems \cite{Pastawski:2015qua,Hayden:2016cfa}. A common feature of these models is that an optimized network that represents a quantum state with particular entanglement resembles slices of holographic geometry. Since this was realized, much work has been dedicated to developing this story by understanding the entanglement and complexity of these holographic networks, as well as to extending the discussion to continuous and strongly interacting CFTs \cite{2012JHEP...10..193N, 2013PhRvL.110j0402H}.

All quantum states can be defined by Feynman path integrals, so the question of extracting holographic geometry from quantum information in CFTs can in principle be posed entirely within this formalism.  Indeed, with the replica trick, the path integral approach to entanglement in CFTs was instrumental in establishing Ryu-Takayanagi formula in AdS/CFT \cite{Lewkowycz:2013nqa}. Similarly, the tensor network observations were generalized into the Euclidean path integral framework in \cite{Miyaji:2016mxg,Caputa:2017urj,Caputa:2017yrh}. It was shown (see section \ref{Review}) that, by a procedure of minimnizing the ``path integral complexity" given by the Liouville action, optimal path integrals in 2d CFTs are performed on hyperbolic geometries (i.e. continuous tensor networks) that can be interpreted as slices of holographic geometries in $AdS_3$ spacetime.

The central role in the path integral construction (both for entanglement as well as path integral optimization) is played by the classical Liouville action, which measures the complexity of the Euclidean path integrals in CFTs. Even though the path integrals are non-unitary, this measure of complexity is the first that can be universally applied to interacting CFTs. It is therefore important to explore it from both a quantum information perspective as well as from holography.

Intuitively, terms in the Liouville action can be compared to an isometry or a disentangler in a MERA \cite{2008PhRvL.101k0501V} tensor network \cite{Czech:2017ryf}. Alternatively, one may interpret the Liouville action (more generally Polyakov action) as Nielsen's geometric complexity for the universal Virasoro circuits \cite{Caputa:2018kdj} built from the energy momentum tensor in 2d CFTs. Moreover, in the framework of \cite{2018arXiv180702501M,2018arXiv180512524M,2018arXiv181200529M}, it is possible to use direct counting arguments supporting the Liouville action as circuit complexity \cite{Camargo:2019isp}. On the gravity side, it was shown \cite{Takayanagi:2018pml} that, for static geometries, the Liouville action is equivalent to the Wheeler-de Witt patch proposal for holographic complexity \cite{Brown:2015lvg}. See \cite{Bhattacharyya:2018wym,2019JHEP...11..132S,PhysRevD.101.026007,Ghodrati:2019bzz,Molina-Vilaplana:2018sfn,Caputa:2018xuf,Erdmenger20} for further developments on the Liouville action in the context of complexity in quantum field theories.

In this work, we develop the idea of path integral optimization and complexity in the framework of inhomogeneous CFTs, obtained by placing a 2d CFT on a position dependent background metric (see below for details). These models play important roles from many body physics to string theory. One of their very interesting features is that by appropriately choosing a background (deforming a CFT) we can tune the structure of entanglement as measured by entanglement entropy. This makes these theories particularly valuable from the perspective of understanding entanglement geometry and complexity using path integrals.

We find that by an appropriate choice of uniformizing coordinates we can solve the optimization problem for inhomogeneous CFTs to find optimal metrics as well as path integral complexity. Optimal metrics are again hyperbolic but have to be supplemented by a position dependent cut-off. Moreover we find an explicit embedding of our solutions into $AdS_3$ geometries for which Einstein's equations are identical to our condition for minimal path integral complexity. We evaluate path integral complexity for vacuum, primary and thermal states and find that the leading divergences are unaffected by inhomogeneiety, but physical scales (e.g. temperatures) are modified in a position dependent manner. Finally we analyze path integral complexity in explicit examples of M\"obius (including SSD), rainbow and constant curvature deformations as well as inhomogeneous CFTs with boundaries.

This paper is organized as follows. In section \ref{Inh} we introduce a class of inhomogeneous CFTs that result from placing a CFT on a curved background metric. In section \ref{PIO} we review the path integral optimization and apply it to inhomogeneous CFTs. In section \ref{Hol} we show how optimal metrics can be mapped onto slices of holographic geometries and in section \ref{KS} we study universal entanglement entropies with associated energy momentum tensors and discuss the connection between kinematic space and our setup. In section \ref{Compl} we evaluate the on-shell Liouville action that computes the path integral complexity in several examples and in section \ref{IBCFT} we discuss boundaries in inhomogeneous BCFTs. Finally we conclude and list some open problems in \ref{Concl} and include more comments and details of the computations in the appendices. 
%%%%%%%%%%%%%%%%%%%%%%%%
%%%%%%%%%%%%%%%%%%%%%%%%
\section{Inhomogeneous CFTs}\label{Inh}
%%%%%%%%%%%%%%%%%%%%%%%%
%%%%%%%%%%%%%%%%%%%%%%%%
While translation invariance is typically assumed when doing quantum field theory, spatial inhomogeneities are ubiquitous in real physical systems. Indeed, the presence of inhomogeneity can lead to novel physics. Static disorder can result in Anderson and many-body localization in lattice systems \cite{PhysRev.109.1492}, defects and boundaries are important for modeling a wide variety of physical phenomena (e.g. \cite{kondo}), and external potentials are vital for performing many experiments. 

With this motivation, some work has been done in recent years to study the effects of placing a 2D conformal field theory on an inhomogeneous background (e.g. \cite{SciPostPhys.2.1.002}). Such a background could result from an external potential or a spatially dependent kinetic energy, and can be encoded in the metric on which the theory lives. More precisely, in a large class of models that have been actively studied up to date, one introduces a background\footnote{In this work we will mostly use the Euclidean signature but the arguments can be discussed naturally with Lorentzian metrics.} 
\begin{equation}
    ds^2= f(x)^2 dt^2 + dx^2,\label{dginh}
\end{equation}
characterized by a single function of the spatial coordinate $f(x)$. Placing a CFT on this metric results in a new ``deformed" Hamiltonian
\be
H[f]=\int dx f(x)h(x),
\ee
where $h(x)$ is the original undeformed Hamiltonian density, proportional to the $tt$-component of the original energy momentum tensor of the CFT. 

In two dimensions, the transformation properties of CFTs under Weyl transformations ensure that we can obtain curved space CFT results with relative ease (see e.g. section \ref{KS}). The physical consequences of such a spatial deformation can be nontrivial. For example, the so-called ``rainbow chain" deformation \cite{2010NJPh...12k3049V,2014JSMTE..10..004R, 2015JSMTE..06..002R,2017JPhA...50p4001R} can result in a ground state with a volume law entanglement entropy, a dramatic violation of the typical $\log \ell$ entanglement entropy with the size on the interval $\ell$ seen in 2D CFTs \cite{2019JSMTE..02.3105A}. 

The sine-squared deformation (SSD) can change the entanglement spectrum of an open boundary condition CFT to that of a CFT with periodic boundary conditions \cite{2009PThPh.122..953G, 2011PhRvB..83f0414H}, and yields a Hamiltonian with nontrivial exactly solvable dynamics \footnote{See \cite{2016IJMPA..3150170I,2016PhRvB..93w5119W, 2018PhRvB..97r4309W, 2018arXiv180500031W, 2019arXiv190805289F,2019JSMTE..03.4001M,Zhang:2019wqo,2019arXiv191207210C} and references therein for applications of inhomogeneous CFTs.}. Despite its many nontrivial features, one can compute many quantities in the SSD model (and other inhomogeneous CFTs) rather easily, owing to the transformation properties of CFTs. One can use the transformation properties of holographic CFTs, for example, to compute the finite temperature entanglement entropy of an arbitrary interval $[x_1,x_2]$ in an SSD deformed CFT on an interval $L$   \cite{2018arXiv181210023M}:
\begin{align}
  S_A(x_1,x_2; \beta)
    =  \frac{c}{3} \log \left[
    \frac{4\beta\sin \left( \frac{x_1}{\pi} \right)
    \sin \left( \frac{x_2}{\pi} \right)}{L \epsilon}
    \
    \sinh \left(
    \frac{L\sin \left(\frac{1}{\pi}(x_2-x_1) \right)}{2 \beta\sin \left( \frac{x_1}{\pi} \right)
    \sin \left( \frac{x_2}{\pi} \right) }
     \right) \right].
\end{align}
One of the characteristic features, common to inhomogeneous theories, is the modification of physical scales (here temperature) in a position dependent manner. Note also that even for the zero temperature case, the entropy looks like that of a theory on a finite circle. This is a particular effect of the SSD deformation \cite{2011PhRvB..83f0414H}.

The above result deviates significantly from the typical, translation-invariant finite temperature CFT entanglement entropy,
\begin{align}
    S_A (x_1,x_2;\beta)= \frac{c}{3} \log
      \left[
      \frac{ \beta}{\pi \epsilon }
      \sinh\left(\frac{\pi (x_2-x_1)}{\beta} \right) 
      \right],
    \label{BTZgeoLen}
\end{align}
and yet can be obtained entirely geometrically, by deforming the background geometry of our CFT. The same can be done for the rainbow chain and any other suitably well-behaved background deformation. All of the CFT formulas can be shown to reproduce the lattice results extremely well \footnote{See \cite{SciPostPhys.2.1.002} for a discussion on the conditions under which the CFT approximation to inhomogeneous systems is valid.}. This result suggests the power and flexibility of using inhomogeneous CFTs, which allow for simple transformation laws and potentially novel physics. This motivates us to use these theories as a testing ground for path integral complexity. Indeed, estimating the complexity of path integrals (see below) that prepare states in inhomogeneous CFTs will allow us compare them to their homogeneous counterparts and analyze the relative difficulty of state preparation that they present.

%In this work, we are also interested in understanding these inhomogeneous CFTs from the perspective of computational complexity. For a given form of the background metric, the Liouville equations of motion become easy to solve, and we can compute the path integral complexity for the theory from the optimized path integral, giving us an estimate of the number of local MERA tensors required to create the ground state of the theory \cite{Caputa:2017yrh,Czech:2017ryf}. Before we do this, we will establish our conventions for working with CFTs on curved backgrounds.

We end this section by reviewing a general argument on the role of the background metric in the ``deformation" of a CFT. Given a stress tensor $T_{\mu \nu}$ and a curved background metric $g_{\mu \nu}$ in $d+1$ dimensions, we can choose a time-slice $\Sigma$ and define the Hamiltonian as the Killing energy:
\begin{equation}
    H= \int_{\Sigma} d^d x \sqrt{h} n_\mu \xi_\nu T^{\mu \nu} \big |_\Sigma,
\end{equation}
where $\sqrt{h}$ is the determinant of the metric induced by $g$ on $\Sigma$, $n_\mu$ is a time-like unit vector normal to $\Sigma$, and $\xi_\nu$ is a time-like Killing vector.% Throughout the paper, we will use 2D Euclidean metrics of the form
%\begin{equation}
 %   ds^2= f(x)^2d \tau^2 + dx^2.\label{dginh}
%\end{equation}

Picking slices of constant $\tau$ in \eqref{dginh}, we have $n^{\mu} \partial_\mu= \frac{1}{f(x)}\partial_\tau$, $\xi^{\mu} \partial_\mu= \partial_\tau$, and $\sqrt{h}=1$. Lowering the indices on these vectors, we have $n_\tau= f(x)$ and $\xi_\tau=f^2(x)$. Meanwhile, the covariant form of the energy density is defined by 
\begin{equation}
    h(x)= T^{\mu \nu} u_\mu u_\nu,
\end{equation}
where $u_\mu$ is the two-velocity of a stationary observer on a slice of constant $\tau$. In our case, this is $u_\mu=n_\mu$, so our energy density becomes
\begin{equation}
    h(x)= T^{\mu \nu} u_\mu u_\nu= T^{\tau \tau}n_\tau n_\tau= T^{\tau \tau} f^2(x).
\end{equation}
Putting all of these pieces together, our Hamiltonian becomes
\begin{equation}
    H= \int_{\tau=const}dx f(x) h(x).
\end{equation}
Clearly, placing a CFT on \eqref{dginh} results in the geometric ``deformation" of a QFT Hamiltonian by the function $f(x)$. This is the starting point and the standard form of geometrically deformed inhomogeneous Hamiltonians that appears in the literature. 

Equivalently, one can consider an inhomogeneous CFT in the Lagrangian formulation. Formally, the action on the curved background \eqref{dginh} can then be seen as that of a CFT deformed by appropriate currents with position dependent couplings (see \ref{appendix:LFS} for the example of a free scalar CFT). 
%%%%%%%%%%%%%%%%%%%%%%%%
%%%%%%%%%%%%%%%%%%%%%%%%
\section{Path Integral optimization for Inhomogeneous CFTs}\label{PIO}
%%%%%%%%%%%%%%%%%%%%%%%%
%%%%%%%%%%%%%%%%%%%%%%%%
In this section, for the purpose of being self-contained, we briefly review the path integral optimization procedure \cite{Caputa:2017urj} for two dimensional CFTs, and later apply it to CFTs on curved backgrounds \eqref{dginh}. More detailed discussion of this approach and generalizations can be found in \cite{Caputa:2017yrh}.
%%%%%%%%%%%%%%%%%%%%%%%%
\subsection{Review}\label{Review}
%%%%%%%%%%%%%%%%%%%%%%%%

The main motivation behind the path integral optimization is the idea that holographic geometry can be read off from CFT states using optimized tensor networks. It was first pointed out by Swingle \cite{2012PhRvD..86f5007S} that tensor networks that efficiently represent CFT states by implementing ``entanglement renormalization" (the so called MERA network \cite{2008PhRvL.101k0501V}), after optimization, resemble a discretization  of a time slice of holographic Anti-de Sitter geometry. This heuristic picture is very appealing for understanding the mechanisms behind AdS/CFT, and much work has been dedicated to implementing it in continuous, interacting holographic models (the continuous MERA tensor network is so far understood only well in free theories). One successful approach has been to start from the most basic object that defines states in all quantum field theories (including strongly interacting ones) --- the Feynman path integral --- and design an optimization procedure that, at the end, allows one to extract slices of holographic geometries. This can be done as follows.

Wave functions in quantum field theories (more appropriately wave functionals) are formally defined by path integrals in Euclidean time $\tau$ and spatial coordinate $x$, performed with boundary conditions at $\tau=\epsilon$ for all of the fields of the theory $\varphi(\epsilon,x)=\tilde{\varphi}(x)$
\begin{equation}
\Psi[\tilde{\varphi}(x)]_{\hat{g}}=\int\left(\prod_x\prod_{\epsilon\le \tau<\infty}D\varphi(\tau,x)\right)\delta\left(\varphi(\epsilon,x)-\tilde{\varphi}(x)\right)e^{-S_E[\varphi,\hat{g}]}.\label{PIDef}
\end{equation}
In the above definition $S_E[\varphi,\hat{g}]$ is the Euclidean action evaluated on a space with background metric $\hat{g}$, which is usually chosen to be flat. Now we want to perform a tensor network-like optimization with this path integral representation as our variational ansatz. The main question that arises is ``which universal parameter should be used in such an optimization?". A natural candidate is the background metric $\hat{g}$ on the Euclidean space on which we perform the path integral. Intuitively, one can think of this metric as a continuous network of tensors that, depending on its form, governs how much path integration should be done in various regions of the Euclidean space to efficiently represent the wave functional with fixed boundary conditions.

More precisely, \cite{Caputa:2017yrh} proposed approximating the path integral \eqref{PIDef} by one with the same boundary condition $\varphi(\epsilon,x)=\tilde{\varphi}(x)$ but computed on a space with a different 2d metric $g$. In 2-dimensions the most general metric can be written as
\begin{equation}
ds^2=g_{\mu\nu}dx^\mu dx^\nu\equiv e^{2\phi}\hat{g}_{\mu\nu}dx^\mu dx^\nu.\label{bgmL}
\end{equation}
The final step is to choose the optimal metric $g$ as the one that minimizes the ``path integral complexity" functional $I[e^{2\phi}\hat{g},\hat{g}]$ of a given wave function that depends on $\phi$ and $\hat{g}$. This functional is defined as the ratio of the two above-defined wave functions with the same boundary condition
\begin{equation}
\frac{\Psi[\tilde{\varphi}(x)]_{e^{2\phi}\hat{g}}}{\Psi[\tilde{\varphi}(x)]_{\hat{g}}}=e^{I[e^{2\phi}\hat{g},\hat{g}]}.\label{ratio}   
\end{equation}
Finally, the actual path integral complexity of a state $\Psi$ is given by the minimal, on-shell value of the functional $I$
\be
C_\Psi\equiv \text{min}_{\phi}I[e^{2\phi}\hat{g},\hat{g}].
\ee
So far the discussion has been general and valid for arbitrary quantum field theories, but in order to have good control of the complexity functional $I$ for the purposes of this work, we now focus on 2d CFTs.

The general metric \eqref{bgmL} is related to the background metric $\hat{g}$ by the Weyl factor $\exp(2\phi)$, and by definition the CFT action is invariant under such a change. However, the CFT path integral measure is not, and is related to the measure on the metric $\hat{g}$ by the exponent of the well-known Liouville action (see e.g. \cite{Seiberg:1990eb}). As a result, the ratio of the two wave functions is universal in 2d CFTs, and the path integral complexity functional is given by
\begin{equation}
I[e^{2\phi}\hat{g},\hat{g}]=S_L[\phi,\hat{g}]-S_L[0,\hat{g}], \label{Ccft}  
\end{equation}
where the Liouville action is given by
\begin{equation}
S_L[\phi,\hat{g}]=\frac{c}{24\pi}\int_{\mathcal{M}} d^2x\sqrt{\hat{g}}\left(\hat{g}^{\mu\nu}\partial_\mu \phi\partial_\nu \phi+e^{2\phi}+\phi R_{\hat{g}}\right)+\frac{c}{12\pi}\int_{\partial\mathcal{M}} ds\sqrt{\hat{h}}(K_{\hat{g}}\phi+\mu_Be^\phi).
\label{liouville_action}
\end{equation}
In the above action $c$ is the central charge of the CFT\footnote{This intuitively corresponds to the bond dimension of the continuous tensor network.}, $R_{\hat{g}}$ is the Ricci scalar of the background metric $\hat{g}$, and we have re-scaled the Liouville field to fix the coefficient of the Liouville exponential potential to $1$. The second part describes the boundary Liouville action with extrinsic curvature $K_{\hat{g}}$ and boundary cosmological constant $\mu_B$. In the complexity functional \eqref{Ccft}, the second part $S_L[0,\hat{g}]$ (required by the transformation of the measure) stands for the $\sqrt{\hat{g}}$ and $\sqrt{\hat{h}}$ in the bulk and boundary actions (that survive from potentials when $\phi=0$). These two subtracted pieces, important for the Liouville as an effective 2d gravity action, give the action $I$ the interpretation of the relative measure of path integral complexity between two metrics $I[g_1,g_2]$. Namely, we have the two important properties
\be
I[g_1,g_2]=-I[g_2,g_1],\qquad I[g_1,g_2]+I[g_2,g_3]=I[g_1,g_3].
\ee
The minimization procedure of the complexity action leads to two constraints for the metrics. In our prescription for 2d CFTs they are the bulk and boundary Liouville equations
\be
R_{\hat{g}}-2\Box_{\hat{g}}\phi+2e^{2\phi}=0=R_g+2,
\ee
and 
\be
K_{\hat{g}}+n^a\partial_a \phi+\mu_B e^{\phi}=0=K_g+\mu_B.
\ee
In the second equality we used the definition of the Ricci scalar and the extrinsic curvature of the full metric $g=e^{2\phi}\hat{g}$, and $\Box_{\hat{g}}$ is the Laplacian of the reference metric $\hat{g}$. The above constraints imply that metrics that optimize the path integral complexity have a constant negative curvature (are hyperbolic). In this work, as in \cite{Caputa:2017yrh}, we will fix the Liouville field at the boundary of the path integration region (see below) and set  $\mu_B=0$ there, but in general $\mu_B$ labels a family of conformal boundary conditions in a BCFT on a finite domain (see \cite{2000hep.th....1012F,2019JHEP...11..132S} and section \ref{IBCFT}).

Generally, complexity $I[g_2,g_1]$ has the ambiguity of the choice of a reference background metric. This can affect the ultimate result. However, in the inhomogeneous case of this work, we will be interested in the comparison between the complexity of a deformed and an undeformed CFT with the same, fixed reference metric. Notice also that on-shell solutions have background curvature and potential related to $\Box \phi$, so the minimal complexity action can be written e.g. only in terms of the potential and $\phi$. 

In \cite{Caputa:2017yrh,Caputa:2017urj} we extensively studied the optimization with the flat reference metric 
\begin{equation}
ds^2=e^{2\phi(z,\bar{z})}dzd\bar{z}.\label{CG}
\end{equation} 
By uniformization, all 2d metrics can be written in this way, and this will play an important role in the optimization for inhomogeneous CFTs below.

For the flat reference metric \eqref{CG} the Liouville equation has the standard form
\begin{equation}
4\partial_z\partial_{\bar{z}}\phi(z,\bar{z})=e^{2\phi(z,\bar{z})},\label{LF}
\end{equation}
and its most general solution \footnote{Modulo symmetries.} is given in terms of two functions $A(z)$ and $B(\bar{z})$ as
\begin{equation}
e^{2\phi(z,\bar{z})}=\frac{4A'(z)B'(\bar{z})}{(1-A(z)B(\bar{z}))^2}.\label{GenSolF}
\end{equation}
In order to ensure that we retain the same cutoff as the original theory, the important part of the path integral optimization is the boundary condition for the Liouville field that has to be fixed such that \cite{Caputa:2017yrh,Caputa:2017urj}
\be
e^{2\phi(z,\bar{z})}|_{bdr}=\frac{1}{\epsilon^2}.\label{bcO}
\ee
For theories with boundaries (BCFTs) the optimization prescription has to be supplemented by the boundary equation (see below and also \cite{2019JHEP...11..132S}).

 We will now discuss/generalize this procedure to CFTs on a curved background metric $\hat{g}$. We will also see how it works in several universal examples and then describe the holographic point of view of the optimal metrics as slices of $AdS_3$ geometries.
%%%%%%%%%%%%%%%%%%%%%%%%%%%%%%%%
\subsection{Inhomogeneous CFTs}
%%%%%%%%%%%%%%%%%%%%%%%%%%%%%%%%
The above algorithm can be naturally generalized to inhomogeneous CFTs. As we reviewed in section \ref{Inh}, the effect of inhomogeneity can be described by introducing a background metric of the form
\begin{equation}
ds^2=\hat{g}_{\mu\nu}dx^{\mu}dx^{\nu}\equiv f(x)^2d\tau^2+dx^2. \label{bgm}
\end{equation}
Therefore, our goal now is to perform an optimization of the path integral that prepares quantum states in a CFT with this fixed background metric. We just need to follow the same steps as above, but take into account the geometry of the underlying space. Indeed, this metric has a non zero curvature given by
\be
R_{\hat{g}}=-\frac{2f''(x)}{f(x)}.\label{Rbg}
\ee
First, we place the inhomogeneous CFT on a more general metric
\begin{equation}
ds^2=e^{2\phi(\tau,x)}(f^2(x)d\tau^2+dx^2),\label{metIH}
\end{equation}
that results in the ratio \eqref{ratio} with Liouville action \eqref{liouville_action} with non-zero Ricci scalar term \eqref{Rbg}.
The optimization (minimization of complexity) is now equivalent to solving the general Liouville equation in a curved background
\begin{equation}
\Box_{\hat{g}}\phi-\frac{1}{2}R_{\hat{g}}=e^{2\phi}, \label{EOM}
\end{equation}
with curvature \eqref{Rbg} and the  Laplace-Beltrami operator of the reference metric
\begin{equation}
\Box_{\hat{g}}=\frac{1}{\sqrt{\hat{g}}}\partial_{\mu}\left(\sqrt{\hat{g}}\hat{g}^{\mu\nu}\partial_\nu\right)=\partial^2_x+\frac{f'}{f}\partial_x+\frac{1}{f^2}\partial^2_\tau.\label{LBO} 
\end{equation}
At first, the task of solving this equation seems much more complicated than the homogeneous case discussed before. However, we can solve the problem with the following ``uniformization" trick. Namely, in two dimensions, we can always find local coordinates in which our metric can be written as \eqref{CG}. In our case, this can be done by first pulling to the the front $f(x)^2$ in \eqref{metIH} and introducing a new variable
\begin{equation}
dy=\frac{dx}{f(x)},\qquad y(x)=\int^x\frac{d\tilde{x}}{f(\tilde{x})},\label{yC}
\end{equation}
as well as $z=\tau+iy$ and $\bar{z}=\tau-iy$, such that we have
\begin{equation}
ds^2=e^{2\phi(t,x)}f^2(x)\left(d\tau^2+\frac{dx^2}{f^2(x)}\right)=e^{2\Phi(z,\bar{z})}dzd\bar{z},
\end{equation}
where in the last step we defined
\begin{equation}
\Phi(z,\bar{z})=\phi(\tau,x)+\log(f(x)).
\end{equation}
Consequently, our equation \eqref{EOM} can be written as the flat space Liouville equation \eqref{LF} for the field $\Phi(z,\bar{z})$. To see this explicitly, note that (suppressing the arguments of $f$)
\begin{equation}
4\partial_z\partial_{\bar{z}}=\partial^2_\tau+\partial^2_y=\partial^2_\tau+ff'\partial_x+f^2\partial^2_x,
\end{equation}
and the Liouville equation \eqref{LF} for $\Phi(z,\bar{z})$ 
%\begin{equation}
%\left[\partial^2_\tau+ff'\partial_x+f^2\partial^2_x\right]\left(\phi(\tau,x)+\log(f)\right)=f^2e^{2\phi(\tau,x)},
%\end{equation}
can be rewritten as
\begin{equation}
\left[\frac{1}{f^2}\partial^2_\tau+\frac{f'}{f}\partial_x+\partial^2_x\right]\phi(\tau,x)+\left[\frac{f'}{f}\partial_x+\partial^2_x\right]\log(f)=e^{2\phi(\tau,x)}.   
\end{equation}
Clearly, the operator on the left is the Laplacian \eqref{LBO} and the first term on the right is minus the curvature \eqref{Rbg} so we recover our equation \eqref{EOM}.

Since we know the general solution \eqref{GenSolF}, the above trick implies that the most general solution of the optimization problem for the inhomogeneous CFTs is given by
\begin{equation}
e^{2\phi(\tau,x)}=\frac{1}{f^2(x)}e^{2\Phi(z,\bar{z})}=\frac{4A'(z)B'(\bar{z})}{f^{2}(x)(1-A(z)B(\bar{z}))^2},
\label{solnCurved}
\end{equation}
where functions $A(z)$ and $B(\bar{z})$ are the same ones that solve the problem of the homogeneous CFT on the same domain (see more below) and the solution depends on $\tau$ and $x$ via $y(x)$ and $z=\tau+iy$ defined above. This is one of our main general results in this section.

Finally, the optimized solution must be supplemented by an appropriate boundary condition. The boundary condition for the field $\phi$ should be chosen such that the solution (our metric) matches the original i.e. ``physical" cut-off as in \eqref{bcO}. From our general solution \eqref{solnCurved} we can already guess that this does not happen at $\tau=\epsilon$ as for homogeneous CFTs. Instead, as we will see in more detail below, this condition forces us to use a position-dependent cut-off at $\tau= \epsilon (x)= \frac{\epsilon}{f(x)}$, such that our boundary condition is satisfied
\begin{equation}
e^{2\phi(\tau,x)}|_{\tau=\frac{\epsilon}{f(x)}}\simeq\frac{1}{\epsilon^2}.
\end{equation}

We will now provide a few universal examples, each of which corresponds to an optimization for a different state in the underlying inhomogeneous CFT. The examples are generalizations of \cite{Caputa:2017yrh} and the comparison of the computations can be made at each step by setting $f(x)=1$.

%%%%%%%%%%%%%%%%%%%%%%%%
%%%%%%%%%%%%%%%%%%%%%%%%
\subsection{Vacuum}
%%%%%%%%%%%%%%%%%%%%%%%%
%%%%%%%%%%%%%%%%%%%%%%%%
The vacuum state of a homogeneous CFT on the real line is usually prepared by a path integral on the flat half plane. Similarly, for the inhomogeneous CFTs we have to perform the path integral on the half plane with metric \eqref{bgm}. This leads to the simplest nontrivial solution of the form (\ref{solnCurved}), which corresponds to the CFT vacuum.

By using the homogeneous solution \cite{Caputa:2017yrh} $A(z)=z$ and $B(\bar{z})=-1/\bar{z}$ such that in the $z=\tau+iy$ coordinates
\begin{equation}
e^{2\Phi(z,\bar{z})}=\frac{4}{(z+\bar{z})^2}=\frac{1}{\tau^2},\label{AdSv}
\end{equation}
for the inhomogeneous CFT on the curved background \eqref{bgm} we have the solution of the Liouville optimization equation
\begin{equation}
e^{2\phi(\tau,x)}=f^{-2}(x)e^{2\Phi(z,\bar{z})}=\frac{1}{f^2(x)\tau^2}.
\end{equation}
As a result, the optimal metric for our path integral is again the hyperbolic metric given by
\be
ds^2=\frac{1}{f^2(x)\tau^2}\left(f^2(x)d\tau^2+dx^2\right)=\frac{d\tau^2}{\tau^2}+\frac{dx^2}{f^2(x)\tau^2}.%=\frac{dt^2+dy^2}{t^2}.
\ee
Introducing the function $f(x)$ effectively stretches or shrinks our spatial $x$ coordinates in a position-dependent manner, and is a manifestation of the particular inhomogeneity that we consider. For $f(x)=1$ (or in the $y$ variable) the metric corresponds to the standard hyperbolic plane $H_2$. 

Another example is the path integral optimization for the vacuum state of a CFT on the circle. In that case, the Euclidean path integral that prepares the state is performed on a disc $|z|\le1$, $z=r e^{ix}$ (with $x\in(0,2\pi)$) and we have a solution for homogeneous CFTs \cite{Caputa:2017yrh} with functions $A(z)=z$ and $B(\bar{z})=\bar{z}$ 
\begin{equation}
e^{2\Phi(z,\bar{z})}=\frac{4}{(1-z\bar{z})^2}=\frac{4}{(1-r^2)^2}.
\end{equation}
To derive the optimized metric for the inhomogeneous CFTs, we can follow the general steps starting from 
\begin{equation}
ds^2=e^{2\phi(r,x)}\left(f(x)^2dr^2+r^2dx^2\right)%=e^{2\phi(r,x)}f^2(x)\left(dr^2+r^2\frac{dx^2}{f(x)^2}\right)
=e^{2\Phi(z,\bar{z})}dzd\bar{z},   
\end{equation}
where now $z=re^{iy}$ and the background metric has a Ricci scalar $R_{\hat{g}}=-\frac{2f''}{r^2f}$ and Laplacian
\be
\Box_{\hat{g}}=\frac{1}{f^2r}\partial_r\left(r\partial_r\right)+\frac{1}{r^2}\left(\frac{f'}{f}\partial_x+\partial^2_x\right).
\ee
From our general solution, we can again see that we get the solution of the Liouville equation
\begin{equation}
e^{2\phi(r,x)}=\frac{4}{f(x)^2(1-r^2)^2},    
\end{equation}
and the optimized metric for the path integral that prepares a vacuum in the inhomogeneous CFT on the circle is given by
\be
ds^2=\frac{4}{(1-r^2)^2}\left(dr^2+r^2\frac{dx^2}{f^2(x)}\right).
\ee 
Again, in this hyperbolic metric, distances in the angular coordinate $x$ are modified by the local function $f(x)$.
%%%%%%%%%%%%%%%%%%%%%%%%
%%%%%%%%%%%%%%%%%%%%%%%%
\subsection{Primary States}
%%%%%%%%%%%%%%%%%%%%%%%%
%%%%%%%%%%%%%%%%%%%%%%%%
Another class of universal solutions to the Liouville equations, the so-called conical defects, can be used in the path integral optimization. These correspond to the CFT eigenstates that are prepared by the insertion of a primary operator of dimension $\Delta=h+\bar{h}$ at the center of the disc \cite{Caputa:2017yrh}. In other words, the path integral that prepares these primary states is again over the unit disc but now with an operator appropriate boundary condition at the center. In coordinates \eqref{CG}, the optimization equation receives an extra contribution from the operator insertion proportional to the $\delta$-function, and takes the form \cite{Caputa:2017yrh}
\be
4\partial_{\bar{z}}\partial_{z}\Phi(z,\bar{z})=e^{2\Phi}+2\pi(\alpha-1)\delta^{(2)}(z),
\ee
and the solution for homogeneous CFTs with $A(z)=z^\alpha$, $B(z)=\bar{z}^{\alpha}$ becomes
\be
e^{2\Phi(z,\bar{z})}=\frac{4\alpha^2}{|z|^{2(1-\alpha)}(1-|z|^{2\alpha})^2}=\frac{4\alpha^2}{r^{2(1-\alpha)}(1-r^{2\alpha})^2}.
\ee
According to our general solution, for inhomogeneous CFTs we will have
\be
e^{2\phi(r,x)}=\frac{4\alpha^2}{f(x)^2r^{2(1-\alpha)}(1-r^{2\alpha})^2},
\ee
and the optimized metric is again hyperbolic with an inhomogeneous function modifying the spatial (angular) direction $x$
\be
ds^2=\frac{4\alpha^2}{r^{2(1-\alpha)}(1-r^{2\alpha})^2}\left(dr^2+r^2\frac{dx^2}{f^2(x)}\right).\label{OptCD}
\ee
However, there is a subtlety here. For the undeformed CFTs we have a relation between the conical deficit parameter $\alpha$ and the dimensions of the operator given by $\alpha=\sqrt{1-\frac{24h}{c}}$. In the inhomogeneous CFTs it is not completely clear to us what the primary operators are (except for the vacuum) and what their spectrum is. In fact, this is still an active area of research even in simple inhomogeneous setups (see e.g. \cite{Katsura:2011ss}). Nevertheless, we still expect that \eqref{OptCD} will be the general optimal metric, but with a more complicated relation between $\alpha$ and the operator dimension. Establishing the precise relation is beyond the scope of this work and we leave it as an open future problem.

Finally, we have to fix the position of the boundary at $r=r_0(x)$ for the correct boundary condition
\be
e^{2\phi(r,x)}|_{r=r_0(x)}\simeq\frac{1}{\epsilon^2}.\label{BCrad}
\ee
This can be done analogously as in \cite{Caputa:2018kdj} followed by a rescaling of the cut-off by $f(x)$. Indeed we can check that the correct relation between the cut-off and $r_0(x)$ is now
\be
\frac{1}{\epsilon}=\frac{2\alpha r^\alpha_0(x)}{f(x)(1-r^{2\alpha}_0(x))},\qquad r^\alpha_0(x)=\sqrt{1+\alpha^2\frac{\epsilon^2}{f^2(x)}}-\alpha\frac{\epsilon}{f(x)},
\ee
and we satisfy the boundary condition \eqref{BCrad}. Setting $\alpha=1$ yields the correct cut-off surface for the vacuum state. This cut-off will play an important role in the computation of the complexity functional in later sections.
%%%%%%%%%%%%%%%%%%%%%%%%
%%%%%%%%%%%%%%%%%%%%%%%%
\subsection{Thermofield double}
%%%%%%%%%%%%%%%%%%%%%%%%
%%%%%%%%%%%%%%%%%%%%%%%%
Finally, we consider the path integral optimization for the thermofield double state, which is prepared by a path integral on a strip of width $\beta/2$. The optimal solution for the homogeneous CFT is found to be \cite{Caputa:2017yrh}
\be
e^{2\Phi(z,\bar{z})}=\frac{4\pi^2}{\beta^2}\frac{1}{\cos^2\left(\frac{\pi(z+\bar{z})}{\beta}\right)},
\ee
parametrized by functions $A(z)=\exp(\frac{2\pi i}{\beta}z)$ and $B(\bar{z})=-\exp(\frac{2\pi i}{\beta}\bar{z})$ where $z=\tau+iy$ and $-\frac{\beta}{4}<\tau<\frac{\beta}{4}$.

Now, for CFTs placed on the background metric \eqref{bgm}, our general solution yields
\be
e^{2\phi(\tau,x)}=\frac{4\pi^2}{\beta^2}\frac{1}{f^2(x)\cos^2\left(\frac{2\pi \tau}{\beta}\right)},
\label{tfd_soln}
\ee
and the optimized hyperbolic metric reads
\be
ds^2=\frac{4\pi^2}{\beta^2}\frac{1}{\cos^2\left(\frac{2\pi \tau}{\beta}\right)}\left(d\tau^2+\frac{dx^2}{f^2(x)}\right).
\ee
We observe the same pattern as in the previous solutions. The optimal metrics are generalizations of the homogeneous solutions obtained by re-scaling the spatial coordinate $x$ by a local function $f(x)$. It is clear from the above examples that at each step we can recover the homogeneous result by setting $f(x)=1$. More complicated solutions can be obtained from these known examples by simply applying appropriate maps to different complex domains.\\
Similarly to the previous solutions, the boundary condition has to be modified to depend on the position. We can see that for the thermofield double solution we have to impose the cut-off at the two boundaries
\be
\tau=\pm \frac{\beta}{4}\mp \frac{\epsilon}{f(x)},\label{COTFD}
\ee
where the exponent of the Liouville factor is fixed to \eqref{bcO}. Again, this will play an important role in the evaluation of the on-shell complexity action later.

In the next section, we will focus on the holographic interpretation of these two-dimensional metrics and, generalizing \cite{Caputa:2017yrh}, we will provide explicit embeddings of the optimal metrics into holographic $AdS_3$ geometries.

%%%%%%%%%%%%%%%%%%%%%%%%
%%%%%%%%%%%%%%%%%%%%%%%%
\section{Optimization and Einstein's equations}\label{Hol}
%%%%%%%%%%%%%%%%%%%%%%%%
%%%%%%%%%%%%%%%%%%%%%%%%
As we reviewed above, one of the goals of the path integral optimization was to extract slices of holographic geometries from path integrals in holographic CFTs. In this light, if we formally identify the Euclidean time $\tau$ with the radial direction of $AdS_3$, then the two-dimensional metrics that optimize path integrals for universal states discussed in \cite{Caputa:2017urj} and above can be seen as constant-time slices of dual $AdS_3$ metrics. This picture is somewhat heuristic, but since it is valid for arbitrary interacting CFTs and generalizes the observations in \cite{2012PhRvD..86f5007S}, it could be a feasible possibility for the emergence of holographic geometries from CFT states.

On the other hand, one can think of holographic $AdS_3$ geometries as a collection of two-dimensional slices on which one can perform path integrals in order to prepare various states in dual CFTs. Or more intuitively, in the language of \cite{Takayanagi:2018pml}, as a collection continuous ``quantum circuits of path-integrations". We will now elaborate on this second picture in our context.

Recall that we can consider a precise slicing of $AdS_3$ space-time in which Einstein's equations are equivalent to solving the Liouville equation \cite{Caputa:2017urj} (i.e. slices of minimial path integral complexity). This is done by the following three-dimensional metric ansatz\footnote{See Appendix \ref{MapsP} for the map to Poincar\'e coordinates.}
\begin{equation}
 ds^2=l^2(d\rho^2+\cosh^2(\rho) e^{2\Phi(z,\bar{z})}dzd\bar{z}).\label{Ansatz3d}
\end{equation}
Indeed, imposing the three-dimensional Einstein equation with negative cosmological constant $\Lambda=-1/l^2$
\be
R_{\mu\nu}-\frac{1}{2}Rg_{\mu\nu}+\Lambda g_{\mu\nu}=0,
\ee
implies that the field $\Phi(z,\bar{z})$ in the ansatz must satisfy the Liouville equation \eqref{LF}, and hence can be parameterized in terms of two functions $A(z)$ and $B(\bar{z})$ as in \eqref{GenSolF}. This way, the optimal metric for the path integral can be identified with a $\rho=0$ slice (generally a constant $\rho$ slice) of the holographic dual $AdS_3$ geometry. This interpretation is also universal (for the universality class of states under consideration) and we do not have to make any identifications between the CFT and bulk coordinates. It is also clear that the above argument can be generalized to higher dimensional path integrals on backgrounds conformally related to flat space.

Let us now see that this line of reasoning neatly generalizes into CFTs on non-trivial background metrics of type \eqref{bgm}. For this, we can simply use the invariance of General Relativity under the choice of coordinates (diff. invariance). Namely, as in the previous sections, we just need to use the uniformization trick and change variables from the ``homogeneous coordinates" to the ``inhomogeneous" ones by\footnote{This change of coordinates can be also used directly in the map from Poincar\'e coordinates explained in appendix \ref{MapsP}}
\begin{equation}
z=\tau+iy,\quad \bar{z}=\tau-iy,\quad dy=\frac{dx}{f(x)}, \label{CoordChange}   
\end{equation}
such that the two-dimensional part of metric \eqref{Ansatz3d} becomes
\begin{equation}
e^{2\Phi(z,\bar{z})}dzd\bar{z}=e^{2\phi(\tau,x)} (f^2(x)d\tau^2+dx^2).
\end{equation}
Replacing this term in our ansatz \eqref{Ansatz3d} and using Einstein's equation again leads to 
\begin{equation}
\Box_{\hat{g}}\phi-\frac{1}{2}R_{\hat{g}}=e^{2\phi},\label{EQGEN}
\end{equation}
which is precisely our condition for minimal path integral complexity for the inhomogeneous CFTs \eqref{EOM}. This way, from the holographic point of view,  the metrics from the path integral optimization in inhomogeneous CFTs can be seen as particular ``inhomogeneous" slices of dual $AdS_3$ geometry at constant $\rho$.

We would like to stress that even though the relation between Einstein's equations and minimization of path integral complexity has been noted before \cite{Caputa:2017yrh,Czech:2017ryf,Caputa:2018kdj}, the diff invariance of the Liouville (or Polyakov) ``complexity action" and its equations of motion are by no means obvious from the quantum mechanical or quantum information perspective. That is why it will be instructive to better understand this property of complexity measures in quantum field theories in the language of gates and operations. Some steps in this direction were initiated in \cite{Caputa:2018kdj} (see also recent work \cite{Camargo:2019isp}) but this program is by no means complete and remains an interesting area for investigation.

Finally, note that the above argument and our anstaz can be generalized to arbitrary background metrics of the form
\be
ds^2=e^{2\phi}\hat{g}_{\mu\nu}dx^\mu dx^\nu.
\ee
Consequently, Einstein's equations with this ansatz are again equivalent to \eqref{EQGEN} with $R_{\hat{g}}$ and $\Box_{\hat{g}}$ computed in metric $\hat{g}$. It may be interesting to use this property to study more general deformations, e.g. the $T\bar{T}$ deformation \cite{Smirnov:2016lqw}, which has been argued to correpsond to a CFT on a random background metric \cite{Cardy:2018sdv,2020arXiv200306300H} (see \cite{PhysRevD.101.026007} for related progress on $T\bar{T}$ deformation and path integral optimization). We leave this as an interesting future problem.

%%%%%%%%%%%%%%%%%%%%%%%%
%%%%%%%%%%%%%%%%%%%%%%%%
\section{Entanglement, Energy and Kinematic Space}\label{KS}
%%%%%%%%%%%%%%%%%%%%%%%%
%%%%%%%%%%%%%%%%%%%%%%%%

Let us now elaborate on entanglement entropy, which can be derived universally for a class of 2d inhomogeneous CFTs \cite{2018arXiv181210023M}. As we discussed before, the inhomogeneous deformations of CFTs that we consider are defined by the background metric
\be
ds^2=f(x)^2d\tau^2+dx^2=f^2(x)\left(d\tau^2+dy^2\right)\equiv e^{2\phi}dzd\bar{z}.
\ee
where we defined the Weyl factor\footnote{different than the Liouville from the path integral optimization!} and uniformizing coordinate
\be
e^{2\phi}=f^2(x),\qquad dy=\frac{dx}{f(x)}.
\ee
In order to compute the entanglement entropy of a single interval in an inhomogeneous CFT,  we will use the transformation property of the two-point functions under Weyl rescaling
\be
\langle \mathcal{O}(z_1,\bar{z}_1)\mathcal{O}(z_2,\bar{z}_2)\rangle_{e^{2\phi}dzd\bar{z}}=\frac{\langle \mathcal{O}(z_1,\bar{z}_1)\mathcal{O}(z_2,\bar{z}_2)\rangle_{dzd\bar{z}}}{e^{\Delta\phi(z_1,\bar{z}_1)}e^{\Delta\phi(z_2,\bar{z}_2)}},
\ee
with $\Delta = h+\bar{h}$, as well as a two-point function\footnote{We use a general two point function that can be obtained by a conformal transformation from the plane} 
\be
\langle \mathcal{O}(z_1,\bar{z}_1)\mathcal{O}(z_2,\bar{z}_2)\rangle=\left(\frac{F(z_1)-F(z_2)}{\epsilon\sqrt{F'(z_1)F'(z_2)}}\right)^{-2h}\left(\frac{\bar{F}(z_1)-\bar{F}(z_2)}{\epsilon\sqrt{\bar{F}'(z_1)\bar{F}'(z_2)}}\right)^{-2\bar{h}}
\ee
where $(F,\bar{F})$ can be thought of as e.g. diff maps between the vacuum and some universal excited states (like descendants of the vacuum, see e.g. \cite{Sheikh-Jabbari:2016znt,Mandal:2014wfa}) of the CFT or further maps to finite temperature or size (to cylinder). %, we can derive a general form of the two point function
%\be
%\langle \mathcal{O}(\tau_1,x_1)\mathcal{O}(\tau_2,x_2)\rangle=\left(\frac{\sqrt{f(x_1)f(x_2)}(F(z_1)-F(z_2))}{\epsilon\sqrt{F'(z_1)F'(z_2)}}\right)^{-2h}\left(\frac{\sqrt{f(x_1)f(x_2)}(\bar{F}(z_1)-\bar{F}(z_2))}{\epsilon\sqrt{\bar{F}'(z_1)\bar{F}'(z_2)}}\right)^{-2\bar{h}}
%\ee
Combining the two expressions, and setting $\tau_i=0$ we then derive the formula for static entanglement entropy of an interval $A=[x_1,x_2]$ in a large class of (geometric) inhomogeneous deformations
\be
S_A(x_1,x_2)=\frac{c}{6}\log\left[\frac{\sqrt{f(x_1)f(x_2)}(F(y(x_1))-F(y(x_2)))}{\epsilon\sqrt{F'(y(x_1))F'(y(x_2))}}\right]+\frac{c}{6}\left(F\leftrightarrow\bar{F}\right),\label{EEd}
\ee
or even more compactly
\be
S_A(x_1,x_2)=\frac{c}{6}\log\left[\frac{F(y(x_2))-F(y(x_1))}{\epsilon\sqrt{\partial_{x_1}F(y(x_1))\partial_{x_2}F(y(x_2))}}\right]+\frac{c}{6}\left(F\leftrightarrow\bar{F}\right).
\ee
For later convenience we can split the above formulas suggestively as
\be
S_A(x_1,x_2)\equiv S^F(x_1,x_2)+S^{\bar{F}}(x_1,x_2).
\ee
The finite temperature SSD example quoted in Section \ref{Inh} corresponds to a particular choice
\be
F(y)=\bar{F}(y)=e^{\frac{2\pi}{\beta}y},\qquad y(x)=-\frac{L}{2\pi}\cot\left(\frac{\pi x}{L}\right).
\ee
In the dual gravitational picture, the inhomogeneous CFTs can be thought to live on appropriately curved foliations of the bulk $AdS_3$ space \cite{2018arXiv181210023M}. Entanglement entropy in the curved space CFTs can then be computed with the Ryu-Takayanagi procedure \cite{Ryu:2006bv} using position-dependent cutoffs determined by the foliation. This formulation is equivalent to the CFT picture above, when $F$ and $y$ are chosen properly.

We can now verify that this general result for entanglement entropy satisfies the Liouville equation. More precisely, we can identify the function in the formula for entanglement entropy as a Liouville field $\phi(x_1,x_2)$
\be
S^F(x_1,x_2)=-\frac{c}{6}\phi(x_1,x_2),\label{EtoL}
\ee
and check that $\phi$ satisfies the Lorentzian Liouville equation\footnote{Analogous arguments go through for $S^{\bar{F}}(x_1,x_2)$.}  \cite{deBoer:2016pqk,Czech:2017ryf}
\be
4\partial_{x_1}\partial_{x_2}\phi=-\frac{4}{\epsilon^2}e^{2\phi}.\label{LLE}
\ee
In terms of the entanglement entropy we then clearly have
\be
\frac{6}{c}\partial_{x_1}\partial_{x_2}S^F=\frac{1}{\epsilon^2}e^{-\frac{12}{c}S^F}.
\ee
One way to interpret this identity is that given the universal entanglement entropy $S(x_1,x_2)=S^F(x_1,x_2)+S^{\bar{F}}(x_1,x_2)$ of a singe interval $A=[x_1,x_2]$ in 2d CFTs, we can associate with it a notion of ``entanglement curvatures" defined as
\be
R^F_{ent}\equiv \frac{48}{c}\,e^{\frac{12}{c}S^F}\,\partial_{x_1}\partial_{x_2}S^F,
\ee
and similarly for $\bar{F}$. Then the Liouville equation can be rewritten as a statement that in 2d CFTs these entanglement curvatures are constant and positive
\be
R^F_{ent}=\frac{8}{\epsilon^2},\qquad R^{\bar{F}}_{ent}=\frac{8}{\epsilon^2}.
\ee
The numerical value of entanglement curvatures may not be important since we can absorb it to the definition of the cut-off $\epsilon$ but the (positive) sign of the curvature is important. Interestingly, from our inhomogeneous context, we can see that these curvatures are invariant under the ``geometric" deformation of the boundary Hamiltonian.

Nevertheless, exploring kinematics of entanglement entropy \eqref{EEd}, we can still extract non-trivial information about stress tensor of the deformed theory. For that we can use the well known fact that the non-linear Liouville equation can be linearlized into a couple of Hill's equations
\be
\left[\partial^2_{x_1}+\frac{1}{2}T_1(x_1)\right]e^{-\phi(x_1,x_2)}=0,\qquad \left[\partial^2_{x_2}+\frac{1}{2}T_2(x_2)\right]e^{-\phi(x_1,x_2)}=0,
\ee
with Liouville stress tensors
\be
T_i(x_i)=2\left(\partial^2_{x_i}\phi-(\partial_{x_i}\phi)^2\right).
\ee
The above equations are just identities involving derivatives of the exponential of $\phi$ but if we now impose the ``chirality" constraints on functions $T(x_i)$
\be
\partial_{x_1}T_2(x_2)=0,\qquad \partial_{x_2}T_1(x_1)=0,
\ee
they are equivalent to the Liouville equation for $\phi$ \eqref{LLE}.

Substituting our identification \eqref{EtoL}, we have the Hill's equations for entanglement entropy of a single interval in a class of inhomogeneous CFTs
\be
\left[\partial^2_{x_1}+\frac{1}{2}T_1(x_1)\right]e^{\frac{6}{c}S^F(x_1,x_2)}=0,\qquad \left[\partial^2_{x_2}+\frac{1}{2}T_2(x_2)\right]e^{\frac{6}{c}S^F(x_1,x_2)}=0,\label{HillEE}
\ee
with
\be
T_{i}(x_i)=-\frac{12}{c}\left(\partial^2_{x_i}S^F+\frac{6}{c}(\partial_{x_i}S^F)^2\right)=-2e^{-\frac{6}{c}S^F}\partial^2_{x_i}e^{\frac{6}{c}S^F}.
\ee
Inserting the explicit form of our solution \eqref{EEd}, yields
\be
T_i(x_i)=\left\{F(y(x_i)),x_i\right\}=\{y(x_i),x_i\}+y'(x_i)^2\{F(y(x_i)),y(x_i)\},\label{EMSD}
\ee
where $\{F,x\}$ is the Schwarzian derivative and $y'(x_i)=1/f(x_i)$. For the entropy of an interval $[x_1,x_2]$ in an SSD deformed CFT on an interval $[0,L]$ at finite temperature, we get
\be
T_i(x_i)=\frac{2 \pi ^2}{L^2}-\frac{\pi ^2}{2 \beta^2 \sin^4\left(\frac{\pi x_i}{L}\right)}.\label{EMSSD}
\ee
We can identify the Liouville stress tensor coming from entanglement entropies of this universal sector of states with the expectation value of the CFT stress tensor itself\footnote{see e.g. \cite{Vaknin:2017yiz,Grumiller:2019xna}} (times a numerical factor of $\frac{c}{12}$). Indeed the functional form of the stress tensor \eqref{EMSD}, as seen from the gravity perspective and Banados geometries matches the interpretation of a stress tensor computed in a CFT on a specific foliation of the bulk (i.e. the curvilinear cut-off advocated in \cite{2018arXiv181210023M}). Moreover, the appearance of the local temperature and explicit position dependence \eqref{EMSSD} further supports this fact in our example. Finally, we can check the consistency with the first law of entanglement entropy \cite{Bhattacharya:2012mi}. Namely, when expanding entanglement entropy for small interval size $x_2=x_1+\l$, $\l\ll 1$ we should get
\be
\Delta S_A\equiv S_A(x_1,x_1+\l)-\frac{c}{3}\log\left(\frac{\l}{\epsilon}\right)= \frac{\pi \l}{3}\Delta E_A+O(l^3)
\ee
where the change of the energy in the small interval $[x_1,x_1+l]$ is given by
\be
\Delta E_A=\int_{\l} dx\, T_{00}\simeq -\frac{1}{2\pi}(T(x_1)+\bar{T}(x_1))\,\l,
\ee
and the proportionality coefficient between the change in the entropy and in the energy is the so-called ``entanglement temperature" (universal, dependent on the shape of the entanglement region). Expanding the entropy in our setup $\eqref{EEd}$ we again derive
\be
\Delta E_A=-\frac{1}{2\pi}\left[\frac{c}{12}\{F(y(x_1)),x_1\}+\frac{c}{12}\{\bar{F}(y(x_1)),x_1\}\right]\l.
\ee
This is clearly consistent with what we obtained by using Hill's equations \eqref{HillEE}.\\

The above observations about entanglement and the Liouville field are closely related to the developments in the Kinematic Space as a 2d geometry with a positive curvature metric 
\be
ds^2=e^{2\phi}dx^1dx^2=\left(\frac{6\epsilon^2}{c}\partial_{x_1}\partial_{x_2} S(x_1,x_2)\right)dx^1dx^2.
\ee
In fact, in the homogeneous CFT case, it was argued  \cite{Czech:2017ryf} that this Liouville field of the kinematic space is closely related to the Liouville field from the path integral optimization\footnote{Roughly analytic continuation and identification of the $(z,\bar{z})$ coordinates with endpoints of the interval}.\\
However, we see that for the inhomogeneous CFTs the path integral optimization and Liouville field are more involved and sensitive to the curved background. In particular, if the relation between the Liouville from path integral optimization and kinematic space was universal, one would expect to obtain a curved kinematic space, but this doesn't seem to occur. Nevertheless, the energy-momentum tensors in the Hill's equations are sensitive to the inhomogeneous background and suggest that there could be a more fine-grained structure associated with kinematic space that goes beyond the Liouville equation. Therefore, the relation between kinematic space and Liouville field from path integral optimization may not be straightforward and it remains to be seen how exactly the two constructions are related.

%%%%%%%%%%%%%%%%%%%%%%%%
%%%%%%%%%%%%%%%%%%%%%%%%
\section{Path Integral Complexity for Inhomogeneous CFTs}\label{Compl}
%%%%%%%%%%%%%%%%%%%%%%%%
%%%%%%%%%%%%%%%%%%%%%%%%
Let us finally discuss the role of the Liouville action as a measure of complexity of continuous tensor network representations of inhomogeneous CFT states. The notion of complexity that we will focus on here is a generalization of the quantum circuit complexity of a tensor network (e.g. MERA or cMERA). Namely, given a tensor network representation of a quantum state that is built from some fixed set of tensors, we can associate with it the notion of complexity as the minimal number of tensors in the optimal network. Heuristically, we think about the path integral as a continuous quantum circuit (though not a unitary one, as we are discussing Euclidean path integrals) and compute its complexity by counting the number of tensors in the circuit after performing the optimization. The natural candidate for counting these continuous tensors is the on-shell Liouville action, which is proportional to the volume of the network.

This intuitive picture can be substantiated by noting that in fact the Liouville (or generally Polyakov) action can be derived as Nielsen's complexity \cite{2005quant.ph..2070N} for the so-called Virasoro circuits defined in \cite{Caputa:2018kdj} built from the CFT energy-momentum tensor operator\footnote{This can be understood as a natural symmetry gate \cite{Magan:2018nmu} in every CFT. See \cite{2019arXiv190803577B} for more discussion.} (see also \cite{Camargo:2019isp} for derivation of the Liouville action by counting tensors in the framework of \cite{2018arXiv180702501M}). Moreover, specifying to free theories, the terms in the Liouville action \eqref{liouville_action} can be intuitively associated with the density of MERA-type tensors \cite{Czech:2017ryf}. Namely, one can argue that the exponential potential term contributes to counting the MERA unitaries and the kinetic term counts the isometric tensors. Following this intuition, the curvature term\footnote{coming from the total curvature $R$ in the Polyakov action} can be interpreted as a contribution to the isometric tensors in a curved background that, depending on the sign, requires using bigger or smaller number of tensors. It could also be thought of as a term that counts tensors of a different type for the MERA network adapted to CFTs in curved backgrounds. 

Despite the similarities and differences with various notions of circuit complexity (See \cite{Jefferson:2017sdb,Chapman:2017rqy,Brown:2019whu,Hackl:2018ptj,Guo:2018kzl,Camargo:2018eof,Liu:2019qyx,Chapman:2018hou,Bhattacharyya:2018bbv,Bhattacharyya:2019kvj,2020arXiv200108664B,Bernamonti:2019zyy,2020arXiv200205779B,Flory:2018akz,Flory:2019kah,Belin:2018bpg,2019arXiv190808514A,2019arXiv190306156A,2020JHEP...01..134B,Balasubramanian:2018hsu,Ali:2018fcz,Ali:2019zcj,Kim:2017qrq,Yang:2018cgx,Yang:2018tpo,Abt:2017pmf,Abt:2018ywl,2020JHEP...01..120D,Erdmenger20} for related some of the recent important developments in circuit complexity in quantum field theories and references therein.), we would like the reader to consider the Liouville action itself as a potentially interesting and independent measure of complexity of path integrals in continuous and interacting CFTs. This is our standpoint and we proceed by exploring this new tool and evaluating the complexity action \eqref{Ccft} with \eqref{liouville_action} on the solutions found in previous sections. We will then draw some conclusions from the perspective of circuit complexity only at the very end.

With a set of solutions to the Liouville equations of motion in hand, we can compute on-shell Liouville actions on curved backgrounds. We will first derive general formulas for the on-shell actions for the vacuum, primary and thermofield-double states for arbitrary functions $f(x)$. We will then consider specific examples of inhomogeneous CFTs, namely the M\"obius and SSD, Rainbow and constant curvature deformations. Our computations closely follow conventions and derivations of \cite{Caputa:2017yrh} for homogeneous CFTs.

%%%%%%%%%%%%%%%%%%%%%%%%
%%%%%%%%%%%%%%%%%%%%%%%%
\subsection{Vacuum and Primaries on the circle}
%%%%%%%%%%%%%%%%%%%%%%%%
%%%%%%%%%%%%%%%%%%%%%%%%
We will start by evaluating the complexity action for both the vacuum and the primaries\footnote{Keeping in mind the subtlety in defining primaries in inhomogeneous CFTs} in an inhomogeneous CFT on a circle \footnote{For simplicity of the discussion we omit the straightforward case of the vacuum on a line and come back to it in the BCFT example.}. The optimized metric is given by
\be
ds^2=e^{2\phi(r,x)}(f^2(x)dr^2+r^2dx^2)\equiv\frac{4\alpha^2}{f^2(x)r^{2(1-\alpha)}(1-r^{2\alpha})^2}(f^2(x)dr^2+r^2dx^2),
\ee
and $\alpha=1$ corresponds to the vacuum. Using the background curvature $R_{\hat{g}}=-\frac{2f''}{r^2f}$ and $\sqrt{\hat{g}}=f(x)r$ we get the bulk part of the Liouville action \eqref{liouville_action}
\bea
S_L[\phi,\hat{g}]&=&\frac{c}{24\pi}\int^{2\pi}_{0}\frac{dx}{f(x)}\int^{r_0(x)}_{\delta}dr\left[\frac{(1-\alpha-(1+\alpha)r^{2\alpha})^2}{r(1-r^{2\alpha})^2}+\frac{4\alpha^2}{r^{1-2\alpha}(1-r^{2\alpha})^2}\right]\nn\\
&+&\frac{c}{24\pi}\int^{2\pi}_{0}dx\left[\frac{f'^2(x)}{f(x)}+2f''(x)\log\left[f(x)\right]\right]\int^{r_0(x)}_{\delta}\frac{dr}{r}\nn\\
&+&\frac{c}{12\pi}\int^{2\pi}_{0}dxf''(x)\int^{r_0(x)}_{\delta}\frac{dr}{r}\log\left[\frac{1}{2\alpha}r^{1-\alpha}(1-r^{2\alpha})\right],
\eea
where, as we discussed in previous sections, for the inhomogeneous CFT solutions we integrate up to the position-dependent cut-off
\be
r=r_0(x)=\left(\sqrt{1+\alpha ^2\frac{\epsilon ^2}{f(x)^2}}-\alpha\frac{\epsilon }{f(x)}\right)^{\frac{1}{\alpha }}.
\ee

Note that we divided the action into the first line, which looks exactly like the bulk contribution for the homogeneous CFTs but with a new, position-dependent cut-off and an overall factor of $1/f(x)$. The remaining terms depend strongly on the deformation (the particular choice of $f(x)$) and vanish for $f(x)=1$.\\

Now we perform integrals over $r$ and expand for small $\epsilon$ and $\delta$
\bea
\int^{r_0(x)}_{\delta}dr\left[\frac{(1-\alpha-(1+\alpha)r^{2\alpha})^2}{r(1-r^{2\alpha})^2}+\frac{4\alpha^2}{r^{1-2\alpha}(1-r^{2\alpha})^2}\right]&\simeq& \frac{2f(x)}{\epsilon}-2\alpha+2\log\left[\frac{2\alpha\epsilon}{f(x)}\right]\nn\\
&-&(1-\alpha)^2\log\delta+O(\epsilon),\label{Int1ep}
\eea
Already from the first term in this integral we can see that the coefficient of the leading $\epsilon$-divergence in complexity will remain the same as for the homogeneous CFTs. Indeed the $f(x)$ in the numerator will cancel with the denominator in the $x$-integral. However, the sub-leading (constant) parts will strongly depend on $f(x)$.

%The remaining integrals are given by
%\be
%\int^{r_0(x)}_{\delta}\frac{dr}{r}\simeq -\log \delta+O(\epsilon),
%\ee
%as well as
%\be
%\int^{r_0(x)}_{\delta}\frac{dr}{r}\log\left[\frac{1}{2\alpha}r^{1-\alpha}(1-r^{2\alpha})\right]\simeq -\frac{\pi^2}{12\alpha}+(1+\log(2\alpha))\log(\delta)+O(\epsilon).
%\ee
In the computation of the complexity functional we also subtract off the volume part that now reads
\be
S_L[0,\hat{g}]=\frac{c}{24\pi}\int d^2x \sqrt{\hat{g}}=\frac{c}{24\pi}\int^{2\pi}_{0}dx f(x)\int^{r_0(x)}_{\delta}dr\,r=\frac{c}{48\pi}\int^{2\pi}_{0}dx f(x)+O(\epsilon).
\ee
Next, we analyze the surface terms. At the $r=r_0(x)$ surface, we have the scaling of our solutions 
\be
\phi(r_0(x),x)\simeq-\log(\epsilon)+O(\epsilon).
\ee
The induced background metric on this boundary is flat $ds^2=dx^2+O(\epsilon)$,
such that the extrinsic curvature on the $r_0(x)$ boundary is
\be
\sqrt{\hat{h}}K_{\hat{g}}|^{r_0(x)}=\frac{1}{f(x)}+O(\epsilon).
\ee
On the other hand, at constant small $r=\delta$, in the inhomogeneous background metric, we have 
\be
\sqrt{\hat{h}}K_{\hat{g}}|_{r=\delta}=\frac{1}{f(x)},
\ee
whereas the value of the Liouville field is given there by
\be
\phi(r=\delta,x)\simeq \log\left(\frac{2\alpha}{f(x)\delta^{1-\alpha}}\right).
\ee
In fact there is an ambiguity in fixing this cut-off and we could have done it in a position-dependent manner e.g. at $r=\delta(x)$. However, as is also done in \cite{Caputa:2017yrh}, we can always add an appropriate ``background charge" at the conical singularity to cancel this divergence (as it is also usually done in the computation of correlation functions from the Liouville action). In any case, we will not be interested in the precise form of the $\delta$-divergent terms for primary operators.

This way we have the boundary contribution (with $\mu_B=0$)
\be
\frac{c}{12\pi}\int_{\partial\mathcal{M}} ds\sqrt{\hat{h}}K_{\hat{g}}\phi=-\frac{c}{12\pi}\int^{2\pi}_{0}\frac{dx}{f(x)}\log\left[\frac{2\alpha\epsilon}{f(x)\delta^{1-\alpha}}\right].
\ee
Note that, except for the $\delta$ term, this expression cancels the logarithmic contribution from \eqref{Int1ep}.

Putting everything together, the on-shell complexity action for an arbitrary $f(x)$ becomes 
\bea
I[e^{2\phi}\hat{g},\hat{g}]&=&\frac{c}{24\pi}\left(\int^{2\pi}_{0}\frac{dx}{f(x)}\left[\frac{2f(x)}{\epsilon}-2\alpha+(1-\alpha^2)\log(\delta)\right]-\frac{1}{2}\int^{2\pi}_{0}dx f(x)\right)\nn\\
&-&\frac{c}{24\pi}\int^{2\pi}_{0}dx\left[\frac{f'^2(x)}{f(x)}+2f''(x)\log(f(x))\right]\log( \delta)\nn\\
&-&\frac{c}{12\pi}\int^{2\pi}_{0}dxf''(x)\left[\frac{\pi^2}{12\alpha}-(1+\log(2\alpha))\log(\delta)\right].\label{OSAV}
\eea
As we mentioned above, it is clear that the leading divergent term, universally (for all $f(x)$), remains the same as in the homogeneous CFTs. However, the second part that depends on the physical scale (dimension of the primary operator) will be proportional to the integral over the coordinate $dx/f(x)$. We can think of this integral as simply the integral over the coordinate $y$ \eqref{yC} that, depending on the choice of $f(x)$, has a different range than the original $x\in(0,2\pi)$. This range enters the complexity action as an overall multiplicative factor. As we will see in the examples, it will lead to the significant changes in path integral complexity when we compare the differences between the homogeneous and inhomogeneous CFTs.

The meaning of this fact from the perspective of complexity is also interesting. Namely, from the perspective of the homogeneous coordinate $x$, the size of the tensor network is now measured by $dx/f(x)$ and, depending on $f(x)$, it can be arbitrarily large. Using the analogy with discrete tensors, we can see that the number of original path integral tensors in the $x$-coordinate needed to prepare our state is significantly different. We will see that this pattern reappears generically for other states below.

The remaining contributions strongly depend on the form of $f(x)$ and boundary terms. The second line in \eqref{OSAV} contains the inhomogeneity term ($x$-derivative from the kinetic term) with $f'$ and the anomaly piece with $f''\log (f)$. This contribution can be integrated by parts to write it as the first term with an additional boundary contribution. The first of them comes from the $x$-derivative in the Liouville kinetic term. If $f(x)$ is such that this term does not vanish it leads to a contribution with logarithmic divergence in $\delta$ from the overall integral over $r$ even for the vacuum with $\alpha=1$. Comparing to the homogeneous CFTs and conical defects from \cite{Caputa:2017yrh} such a divergence indicates that the state under consideration is excited. This may be not surprising, since in two dimensions excitations can be equivalently generated by a change in the cut-off.

%%%%%%%%%%%%%%%%%%%%%%%%
%%%%%%%%%%%%%%%%%%%%%%%%
\subsection{Thermofield double}
%%%%%%%%%%%%%%%%%%%%%%%%
%%%%%%%%%%%%%%%%%%%%%%%%
In this section we perform a similar analysis for the thermofield-double states in inhomogeneous CFTs. Using the solution \eqref{tfd_soln} with an appropriate cut-off \eqref{COTFD} we can evaluate the bulk part of the on-shell Liouville action
\begin{eqnarray}
S_L[\phi,\hat{g}]&=&\frac{c}{24\pi}\left(\frac{2\pi}{\beta}\right)^2\int^{2\pi}_0 \frac{dx}{f(x)}\int^{\frac{\beta}{4}-\frac{\epsilon}{f(x)}}_{-\frac{\beta}{4}+\frac{\epsilon}{f(x)}}d\tau \left(\tan^2\frac{2\pi \tau}{\beta}+\frac{1}{\cos^2\frac{2\pi \tau}{\beta}}\right)\nn\\
&+&\frac{c}{24\pi}\int^{2\pi}_0dx \left(\frac{f'^2(x)}{f(x)}+2f''(x)\log f(x)\right)\left(\frac{\beta}{2}+O(\epsilon)\right)\nn\\
&+&\frac{c}{12\pi}\int^{2\pi}_0 dx f''(x)\int^{\frac{\beta}{4}-\frac{\epsilon}{f(x)}}_{-\frac{\beta}{4}+\frac{\epsilon}{f(x)}}d\tau \log\left[\frac{\beta}{2\pi} \cos\frac{2\pi \tau}{\beta}\right].
\end{eqnarray}
The integrals over $\tau$ are computed and expanded in $\epsilon$ to
\be
\int^{\frac{\beta}{4}-\frac{\epsilon}{f(x)}}_{-\frac{\beta}{4}+\frac{\epsilon}{f(x)}}d\tau \left(\tan^2\frac{2\pi \tau}{\beta}+\frac{1}{\cos^2\frac{2\pi \tau}{\beta}}\right)=\frac{\beta^2}{\pi^2}\left(\frac{f(x)}{\epsilon}-\frac{\pi^2}{2\beta}+O(\epsilon)\right),
\ee
Notice that we again have the same pattern as in the previous section. Namely, the leading divergence is multiplied by a factor of $f(x)$ that will cancel the denominator of the $x$-integral and leave the leading divergence of the path integral complexity unchanged.
%We also have
%\be
%\int^{\frac{\beta}{4}-\frac{\epsilon}{f(x)}}_{-\frac{\beta}{4}+\frac{\epsilon}{f(x)}}d\tau \log\left[\frac{\beta}{2\pi} \cos\frac{2\pi \tau}{\beta}\right]=\frac{\beta}{2}\log\left(\frac{\beta}{4\pi}\right)+O(\epsilon),
%\ee
%and in the complexity functional we subtract the volume term that is given now by
%\be
%S_L[0,\hat{g}]=\frac{c}{24\pi}\int d^2x \sqrt{\hat{g}}=\frac{c}{24\pi}\frac{\beta}{2}\int^{2\pi}_{0}dx f(x)+O(\epsilon).
%\ee
Regarding the boundary contribution, at both boundaries, the induced metric is flat $ds^2=dx^2+O(\epsilon^2)$ and the extrinsic curvatures are of order $O(\epsilon)$.

Putting everything together, the on-shell complexity functional of the thermofield double state for arbitrary $f(x)$ in inhomogeneous CFTs becomes
\begin{eqnarray}
I[e^{2\phi}\hat{g},\hat{g}]&=&\frac{c}{3}\left(\frac{1}{\epsilon}-\frac{\pi}{4\beta}\int^{2\pi}_0 \frac{dx}{f(x)}\right)-\frac{c}{24\pi}\frac{\beta}{2}\int^{2\pi}_0dx\,f(x)\nn\\
&+&\frac{c\beta}{48\pi}\int^{2\pi}_0dx \left(\frac{f'^2(x)}{f(x)}+2f''(x)\log f(x)\right)\nn\\
&+&\frac{c\beta}{24\pi}\log\left(\frac{\beta}{4\pi}\right)\int^{2\pi}_0 dx f''(x)+O(\epsilon).
\label{tfdint}
\end{eqnarray}
This expression shows analogous features to the primary states in the previous section. The first line is the generalization of the homogeneous result from \cite{Caputa:2017yrh} with the same leading divergence. Interestingly, the physical length-scale $\beta$ (or the inverse temperature) is again multiplied by the integral over the $y$-coordinate. This can be thought of as a new effective temperature induced by inhomogeneity. Moreover, this term is also the only $f$-dependent term that doesn't vanish at infinite temperature. 

The second and third line are new, inhomogeneous, contributions from the kinetic term ($x$-derivative) and the Ricci scalar of the background. They are strongly dependent on the choice of $f(x)$, and in the next section we will analyze explicit examples to better understand the effect of these contributions. 

%%%%%%%%%%%%%%%%%%%%%%%%%%%%
\subsection{Examples}
%%%%%%%%%%%%%%%%%%%%%%%%%%%%
In this section we evaluate a few concrete examples to see the effect of inhomogeneity on path integral complexity in concrete deformations.

\subsubsection{M\"obius and Sine-Squared Deformation (SSD)}

We start by computing the path integral complexity for the class of deformations known as the M\"obius deformations, which can be obtained by taking the deformed Hamiltonian to be a linear combination of the $L_1$, $L_{-1}$, and $L_0$ Virasoro generators. For a strip of width $L$, the associated envelope function is \cite{2016arXiv160309543O}
\bea
f(x)=1-\tanh(2\gamma)\cos\frac{2\pi x}{L}.\label{MdF}
\eea
An important limit of this deformation occurs when we take $\gamma \rightarrow \infty$, in which case we obtain the sine-squared deformation (SSD) \cite{2016arXiv160309543O}:
\bea
f(x) \rightarrow 2 \sin ^2 \frac{\pi x}{L}.
\eea

M\"obius deformations and the SSD possess many interesting features, which can be found throughout the condensed matter and string theory literature. In particular, the SSD seems to map the entanglement spectrum of an open boundary condition CFT to that of a periodic boundary condition CFT by suppressing edge effects. This highly nontrivial property suggests that its complexity may have interesting features. 

Here we compute the on-shell Liouville action for the vacuum on a circle with a M\"obius deformed background (and therefore setting $L=2 \pi$). %We refer the reader to Appendix \ref{strip_soln} for the solution to the Liouville equations of motion on a finite strip --- the setting in which the
The M\"obius deformation is usually studied at finite size, but here we use the disk rather than the strip for simplicity, and the results are interesting nevertheless.

Evaluating the integrals for the M\"obius deformation function \eqref{MdF}
\be
\int^{2\pi}_0\frac{dx}{f(x)}=2\pi\cosh(2\gamma),\qquad \int^{2\pi}_0 f(x)dx=2\pi,\qquad\int^{2\pi}_0 f''(x)dx=0,\label{IntfM}
\ee
and 
\bea
\int^{2\pi}_0dx \left(\frac{f'^2}{f}+2f''\log f(x)\right)=-2\pi\tanh(\gamma)\tanh(2\gamma).
\eea
Interestingly, the background curvatures for the disk and strip (TFD) domains respectively are
\be
R^{disc}_{\hat{g}}=\frac{1}{r^2}R^{TFD}_{\hat{g}}=-\frac{1}{r^2}\frac{2 \tanh (2 \gamma ) \cos (x)}{1-\tanh (2 \gamma ) \cos (x)}\label{RMobi}
\ee
and for $x\in[0,2\pi]$ they are negative for $x\le \frac{\pi}{2}$ and $x\ge \frac{3\pi}{2}$ and positive in between. Therefore, our optimal continuous tensor network, with negative Ricci Scalar (Liouville e.o.m.), can still lead to optimal path integrals for CFTs on backgrounds with locally positive curvatures. It would be very interesting to understand this phenomenon from the perspective of other CFT tensor networks like (c)MERA and we hope to come back to this problem in the future. 

Putting together all of the pieces, we get the relative path integral complexity for the class of primary states (that we denote $I^{(\alpha)}$) in M\"obius deformed CFTs is given by
\bea
I^{(\alpha)}[e^{2\phi}\hat{g},\hat{g}]&=&\frac{c}{6}\left(\frac{1}{\epsilon}-\frac{1}{2} \left[(\alpha^2-1)\log(\delta)+2\alpha\right]\cosh(2\gamma)-\frac{1}{4}\right)\nn\\
&+&\frac{c}{12}\tanh(\gamma)\tanh(2\gamma)\log (\delta)+O(\epsilon).
\eea

Similarly, the path integral complexity for the thermofield double in M\"obius  inhomogeneous CFTs becomes
\begin{eqnarray}
I^{(TFD)}[e^{2\phi}\hat{g},\hat{g}]&=&\frac{c}{3}\left(\frac{1}{\epsilon}-\frac{\pi^2}{2\beta}\cosh(2\gamma)-\frac{\beta}{8}\right)
-\frac{c\beta}{24}\tanh(\gamma)\tanh(2\gamma)+O(\epsilon).\label{MobTFD}\nn\\
\end{eqnarray}
Both expressions reduce to homogeneous results \cite{Caputa:2017yrh} for $\gamma=0$ and in each of them the leading divergences are unaffected. For large $\gamma$ (closer to the SSD deformation) the $\cosh(2\gamma)$ grows exponentially\footnote{One could in fact do the SSD computation independently and notice that the first integral in \eqref{IntfM} diverges on the domain $[0,2\pi]$. It can then be regulated by a small cut-off $\delta$ (i.e. evaluated on $[\delta,2\pi-\delta]$) and direct comparison with the limit of the Mobius result \eqref{MobTFD} gives the relation $\frac{4}{\delta}=\pi e^{2\gamma}$ as $\gamma\to\infty$.} and leads to a strong decrease (relative to the infinite leading part) in path integral complexity. On the other hand the $\tanh(\gamma)$ terms become unity and lead to constant negative contributions for SSD. 

As long as the parts with $\log(\delta)$ are ambiguous and in principle could be removed by addition of background charges, the $\alpha$ term for the primaries and $1/\beta$ term for the thermofield double strongly confirm that the inhomogeneous deformations decrease the path integral complexity. From the perspective of the Liouville action that, evaluated on-shell, computes the volume of the continuous tensor network needed to prepare a state, it is natural that the position dependent modification of the cut-off reduces the volume and number of tensors needed. On the other hand, one could expect that the network with non-trivial background curvature \eqref{RMobi}, would be more complex. The above results then provide sharp, interesting questions that we will only be able to address once we better understand the exact correspondence between the Liouville metrics and continuous tensor networks, as well as the nature of complexity as defined by the Liouville action.

As for the leading divergence, its universality and indifference to the geometric deformation may be related to the fact that both ``continuous tensor networks" have constant total negative curvature $R$. It is also instructive to compare the changes in the path integral complexity due to the inhomogeneous deformation. %In gate-based language, this amounts to ``changing the reference state" from which we measure complexity. 
We define $\Delta I$
\be
\Delta I_f\equiv I[e^{2\phi_f}\hat{g}_f,\hat{g}_f]-I[e^{2\phi}\hat{g},\hat{g}],
\ee
as the difference between the path integral complexity computed for a particular state (of a fixed domain) in an inhomogeneous CFT (with $\phi$ and $\hat{g}$ containing $f(x)$) and the path integral complexity of this state in a homogeneous CFT. %We then have for the primary states labeled by $\alpha$
%\bea
%\Delta I^{(\alpha)}_f&=&- \left(\frac{c}{12}(\alpha^2-1)\log(\delta)+\frac{c}{3}\alpha\right)(\cosh(2\gamma)-1)\nn\\
%&+&\frac{c}{12}\tanh(\gamma)\tanh(2\gamma)\log (\delta)+O(\epsilon).
%\eea
For example, for the thermofield double we obtain
\begin{eqnarray}
\Delta I^{(TFD)}_f=-\frac{c\pi^2}{6\beta}(\cosh(2\gamma)-1)
-\frac{c\beta}{24}\tanh(\gamma)\tanh(2\gamma)+O(\epsilon).
\end{eqnarray}
If we expand this result for small deformation parameter $\gamma$, the leading contribution comes at second order in $\gamma$
\be
\Delta I^{(TFD)}_f=-2\left(\frac{c\pi^2}{6\beta}+\frac{c\beta}{24}\right)\gamma^2+O(\gamma^4).
\ee
Moreover, the coefficient of $\gamma^2$ is (twice) the finite part of the undeformed answer for the TFD path integral complexity. Naively, by varying the action with respect to small changes in the metric $g^{\tau\tau}=\delta f$ we would expect to simply get the integral over the $T_{\tau\tau}\delta g^{\tau\tau}$ in the leading answer but we should be more careful now since also the integral over $\tau$ depends on $\delta f$. It would be interesting to perform a more careful analysis of the infinitesimal variations with boundaries and verify if the above behaviour of $\Delta I$ is universal (i.e. related to the first law of complexity observed in \cite{Bernamonti:2019zyy}) or rather a property of M\"obius deformations. 

%%%%%%%%%%%%%%%%%%%%%%%%%%%%%%%
\subsubsection{Rainbow Chain}
%%%%%%%%%%%%%%%%%%%%%%%%%%%%%%%

We now turn to another interesting example of a CFT on a curved background --- the so-called ``rainbow chain". Deforming our geometry to the following metric
\bea
ds^2= e^{-2h|x|}d\tau^2 + dx^2,\qquad f(x)=e^{-h|x|},
\eea
with background Ricci scalar
\be
R_{\hat{g}}=-\frac{2f''(x)}{f(x)}=-2h^2+4h\delta(x),
\ee
hence the CFT is placed on a negatively curved background with a delta function at the origin (in the position of a defect).

Such a background leads to a ground state with distinct violations of the typical $\log \ell$ entanglement. Subintervals that do not include the origin demonstrate volume law entanglement scaling, while intervals symmetric about the origin --- or defect --- show an area law \cite{2018arXiv181210023M}. In a spin chain this ground state consists of a collection of concentric Bell pairs, symmetric about the defect --- hence the ``rainbow". Further properties of the rainbow chain have been explored in the literature. The highly nonlocal difference in the rainbow ground state from the typical CFT vacuum makes it a particularly interesting candidate to explore with path integral optimization.

The optimized Liouville field for the vacuum on the rainbow background is given by
\be
e^{2\phi(\tau,x)}=\frac{1}{\tau^2f(x)^2}=\frac{1}{\tau^2}e^{2h|x|},
\ee
and also
\be
\int^{L}_{-L}\frac{dx}{f(x)}=\frac{2}{h}\left(e^{hL}-1\right),\quad \int^{L}_{-L}f(x)dx=\frac{2}{h}\left(1-e^{-hL}\right),\quad\int^{L}_{-L} f''(x)dx=-2h\,e^{-hL},
\ee
and 
\bea
\int^{L}_{-L}dx \left(\frac{f'^2(x)}{f(x)}+2f''(x)\log f(x)\right)=2h(1+2hL)e^{-hL}.
\eea
%we have $\sqrt{\hat{g}}=f(x)$ and from
%\bea
%\sqrt{\hat{g}}(\hat{g}^{\mu\nu}\partial_\mu\phi\partial_\nu\phi+e^{2\phi})&=&\frac{2}{\tau^2f(x)}+\frac{f'^2(x)}{f(x)},\nn\\
%\sqrt{\hat{g}}R_{\hat{g}}\phi&=&2f''(x)\log\left(\tau f(x)\right),
%\eea
%We can then evaluate the bulk part of the Liouville action
%\bea
 %S_L[\phi, \hat{g}]&=&\frac{c}{24\pi}\int_{-x_\infty}^{x_\infty} dx \int^{\tau_\infty}_{\frac{\epsilon}{f(x)}} d\tau\left[\frac{2}{\tau^2f(x)}+\frac{f'^2(x)}{f(x)}+2f''(x)\log\left(\tau f(x)\right)\right],\nn\\
 %&=&\frac{c}{24\pi}\int_{-x_\infty}^{x_\infty} dx \left[\frac{2}{\epsilon}+\frac{\tau_\infty f'(x)^2}{f(x)}+2 \tau_\infty  f''(x) (\log (\tau_\infty  f(x))-1)\right]+O(\epsilon),\nn\\
 %&=&\frac{c}{24\pi} \left[\frac{4x_\infty}{\epsilon}+\tau_\infty\int_{-x_\infty}^{x_\infty} dx\left(\frac{ f'(x)^2}{f(x)}+2f''(x)\log(f(x))\right)\right],\nn\\
 %&+&\frac{c}{12\pi} \tau_\infty (\log (\tau_\infty)-1)\int_{-x_\infty}^{x_\infty} dx f''(x)+O(\epsilon),\nn\\
 %&=&\frac{c}{6\pi}\frac{x_\infty}{\epsilon}+O(\epsilon)
%\eea
% So this doesn't look like an interesting example. \textcolor{red}{Perhaps we should make a note about the order of limits here} \\
We can then evaluate the complexity for e.g. the TFD in Rainbow deformed CFT (taking $x\in[-L,L]$). That becomes 
\begin{eqnarray}
I[e^{2\phi}\hat{g},\hat{g}]&=&\frac{c}{3}\left(\frac{1}{\epsilon}-\frac{\pi}{2\beta h}\left(e^{hL}-1\right)\right)-\frac{c}{24\pi}\frac{\beta}{h}\left(1-e^{-hL}\right)\nn\\
&+&\frac{c\beta h}{24\pi}(1+2hL)e^{-hL}-\frac{c\beta h}{12\pi}\log\left(\frac{\beta}{4\pi}\right)\,e^{-hL}+O(\epsilon).
\end{eqnarray}
Interestingly, we can see that physical scales are modified by the curvature $h$ of the background. We will now analyze a simpler case of constant background curvature to see if the finite temperature complexity indeed scales with curvature in a universal way.

%%%%%%%%%%%%%%%%%%%%%%%%%%%%%%%
\subsubsection{Constant Curvature deformations}
%%%%%%%%%%%%%%%%%%%%%%%%%%%%%%%
Starting with a general background metric \eqref{bgm} we can consider a special class of inhomogeneous deformations $f(x)$ with constant background curvature 
\be
R_{\hat{g}}=-\frac{2f''(x)}{f(x)}=-2\kappa.
\ee
For example, we can consider 
\be
f(x)=\sin(\sqrt{\kappa}x).
\ee
which is periodic and yields constant positive curvature $\kappa$. Now as an example we evaluate the on shell path integral complexity (\ref{tfdint}) for the thermofield double on this background on $x\in [0,\frac{\pi}{\sqrt{\kappa}}]$ to obtain

\be
I\left[e^{2\phi} \hat{g},\hat{g} \right]= \frac{c}{3} \left( \frac{1}{\epsilon} +\frac{\pi}{2 \beta \sqrt{\kappa}} \log \left( \frac{\epsilon}{2}\right)  + \frac{\beta \sqrt{\kappa}}{8 \pi } \log 2\epsilon   \right) + \mathcal{O}(1).
\ee

Note that we take $x\in \left [0 +\frac{\epsilon}{\sqrt{\kappa}}, \frac{\pi-\epsilon}{\sqrt{\kappa}} \right ]$  in order to introduce a spatial cutoff $\epsilon$. It is interesting to note that increasing the curvature has the effect of decreasing the effective temperature. However, unlike the M\"obius case, the $\beta$-dependent terms \textit{add} to the complexity, rather than decrease it. This can be naturally interpreted as the local geometry of the tensor network that, depending on the sign of the curvature, requires the use of smaller or larger numbers of tensors in order to prepare a quantum state using path integrals.

%we may need to make corrections to our intuition about curved cutoffs decreasing the volume of the path integral manifold. % Clearly the relation between background curvature and path integral complexity is not so trivial, and should be investigated in further detail.

%%%%%%%%%%%%%%%%%%%%%%%%%%%%
\section{Boundaries in inhomogeneous CFTs}\label{IBCFT}
%%%%%%%%%%%%%%%%%%%%%%%%%%%%
In the above discussions, in order to understand universal features of complexity in inhomogeneous setups, we did not pay careful attention to the physical boundaries. More precisely, when evaluating the complexity action on a finite spatial domain (say an interval in the rainbow or constant curvature deformations) we only performed the integrals on the original domain, without taking care of the new optimized shape of the boundary region. Therefore, in this section, we finish our discussion by coming back to this issue.

As we reviewed in section (\ref{Review}), path integral optimization in the context of a boundary CFT (BCFT) has to be supplemented by the Neumann condition on the optimal boundary surface\footnote{This was used in \cite{Caputa:2017yrh} to show that optimization for density matrices leads to (two copies of) the entanglement wedge. See also \cite{2019JHEP...11..132S} for careful analysis of BCFT.} 
\be
K^{(i)}_g+\mu^{(i)}_B=0=K^{(i)}_{\hat{g}}+(n^{(i)})^a\partial_a \phi+\mu^{(i)}_B e^{\phi}\label{BCIN}
\ee
on each of the physical boundaries labeled by $(i)$ \footnote{Note that this is not the case for the cut-off surface of the Euclidean path integral where we simply fix $\exp{2\phi}|_{bdr}=1/\epsilon^2$}. Solving this condition yields a new ``optimal shape of the boundary surface" that we then have to include in the computation of the complexity functional. Let us see explicitly how this works for the vacuum of an inhomogeneous CFT on a strip $x\in [-L,L]$.

We start again with the inhomogeneous solution on the upper half plane
\be
ds^2=e^{2\phi(\tau,x)}(f^2(x)d\tau^2+dx^2)=\frac{1}{\tau^2f^2(x)}(f^2(x)d\tau^2+dx^2),
\ee
that is integrated from $\tau=\epsilon/f(x)$ where $\phi=-\log (\epsilon)$ up to $\tau_\infty\to\infty$. Then, we limit ourselves to $x\in[-L,L]$ and, using \eqref{BCIN}, we want to determine the shape of the two boundary surfaces (treated both at once) $\tau=F_{\pm}(x)$ that start at $-L$ and $L$, respectively. This is done as follows.

We can first compute the unit normal vectors 
\be
(n^\pm)_a=s\frac{f(x)}{\sqrt{1+f^2(x)F'^2_{\pm}}}\{1,-F'_\pm(x)\},
\ee
with $s=\pm 1$, and the background extrinsic curvatures 
\be
K^{(\pm)}_{\hat{g}}=-s\frac{f'(x) F'_\pm(x) \left(2+f(x)^2 F'_\pm(x)^2\right)+f(x) F_\pm''(x)}{\left(1+f(x)^2 F_\pm'(x)^2\right)^{3/2}}.
\ee
Then the derivatives of the Liouville field in the normal directions are
\be
(n^\pm)^a\partial_a\phi=s\frac{f(x) f'(x)  F_\pm(x)F_\pm'(x)-1}{f(x) F_\pm(x) \sqrt{1+f(x)^2 F_\pm'(x)^2}}.
\ee
Putting this together into the condition for the shape of the physical boundary \eqref{BCIN}, we can write it as
\be
\mu^{(\pm)}_B=s\frac{1+f(x)f'(x)  F_\pm(x) F_\pm'(x)+f(x)^2 \left(F_\pm'(x)^2+F_\pm(x)
   F_\pm''(x)\right)}{\left(1+f(x)^2 F_\pm'(x)^2\right)^{3/2}}.
\ee
Now recall that for homogeneous CFTs, the constant $\mu_B$ labels a family of conformal boundary conditions \cite{2000hep.th....1012F}. At first, in inhomogeneous CFTs, the above $\mu_B$ looks more complicated, but it is natural to expect that this parameter should remain constant\footnote{Despite the feature that other scales in inhomogeneous CFTs are modified in a position dependent way.}. Indeed, we can satisfy the condition with the following ansatz for the boundary curves
\be
F_\pm(x)=\alpha_\pm y(x)+b_\pm,
\ee
where the function $y(x)$ is the coordinate that we used to uniformize our backgrounds, namely $y'(x)=\frac{1}{f(x)}$ and $\alpha_\pm$ and $\beta_\pm$ are constants. This way, we have
\be
F'_\pm(x)=\frac{\alpha_\pm}{f(x)},\qquad F''_\pm(x)=-\frac{\alpha_\pm f'(x)}{f^2(x)},\label{FFP}
\ee
and our boundary constraint reduces to that of the homogeneous CFT \cite{2019JHEP...11..132S}\footnote{Even though the relation between $\alpha$ and $\mu_B$ remains the same the identification between the inhomogeneous CFT data may be much more complicated than in homogeneous setups (say RCFTs).} 
\be
\mu^{(\pm)}_B=s\frac{1}{\sqrt{1+\alpha^2_\pm}}.
\ee
We can then find the coefficients of the boundary curves
\be
\alpha_\pm=\mp\sqrt{\frac{1-(\mu^{(\pm)})^2_B}{(\mu^{(\pm)})^2_B}}\equiv \mp \alpha.
\ee
Note that we have a sign ambiguity in the above expressions. Firstly, the sign of $\mu_B$ is arbitrary in our context but, in the Liouville theory, can be fixed to $0<\mu_B<1$ \cite{2000hep.th....1012F} by additional quantum constraints on the correlation functions (see also discussion in \cite{2019JHEP...11..132S}). We expect that this range becomes more important in the evaluation of complexity in a BCFT with operator insertions (defects). The constraint also allows for two signs of $\alpha$. From the perspective of the boundary, in the homogeneous case, the signs will correspond to two boundaries opening up the strip or closing it by moving towards each other. However, from the perspective of the function $f(x)$ that is only defined on the physical region $x\in [-L,L]$, in order to integrate it in the Liouville action with $\tau >\epsilon/f(x)$, only one choice is allowed. Namely, after the optimization, the two boundaries have to move towards each other (see example below).

Finally, we can fix the constants $b_\pm$ by matching with the boundary condition for the cut-off surface at $\tau=\epsilon/f(x)$ in the physical region. This way we obtain the two curves
\bea
F_-(x)&=&\alpha\left[y(x)-y(-L+\epsilon)\right]+\frac{\epsilon}{f(-L+\epsilon)},\\
F_+(x)&=&-\alpha \left[y(x)-y(L-\epsilon)\right]+\frac{\epsilon}{f(L-\epsilon)},\label{Fbcft}
\eea
where
\be
y(x)=\int^x \frac{dx'}{f(x')}.
\ee
Notice that, depending on $f(x)$, boundaries can be very complicated functions of the physical position $x$. However, in the coordinate $y$ they become straight lines.

%%%%%%%%%%%%%%%%%%%%%%%%%%%%%%%
\subsection{Application to SSD}
%%%%%%%%%%%%%%%%%%%%%%%%%%%%%%%
As an interesting example of the boundary phenomena, consider the SSD deformed CFT. One of the defining features of this deformation is that entanglement entropy of an SSD deformed model on a strip with open boundary conditions has the same form as entropy on a system with periodic boundary conditions \cite{2011PhRvB..83f0414H}. It is therefore interesting to see if or how this property can be reflected in the path integral geometry and complexity.

More precisely, we will start with the strip $x\in[-L,L]$ where the SSD deformation has the form (CSD)
\be
f(x)=2\cos^2\left(\frac{\pi x}{2L}\right),\label{fSSDs}
\ee
and we also get the $y(x)$ coordinate  
\be
y(x)=\frac{L}{\pi}\tan \left(\frac{\pi  x}{2 L}\right).\label{YSSD}
\ee
If we first naively evaluate the bulk complexity functional neglecting the boundary terms, we get% \textcolor{red}{(I get a different result than you so can you double check?)}
\be
S_L=\frac{c}{12\pi}\left[\frac{2L}{\epsilon}-\frac{\pi^2}{L}\tau_\infty-4\log(\epsilon)+O(\epsilon)\right],
\ee
again the leading divergence is unaffected and we get a divergent contribution $\tau_\infty$ from the infinite strip region. This naive computation clearly yields a different result than expected form the behaviour of entanglement entropy.
\begin{figure}
    \centering
    \includegraphics[width=9cm]{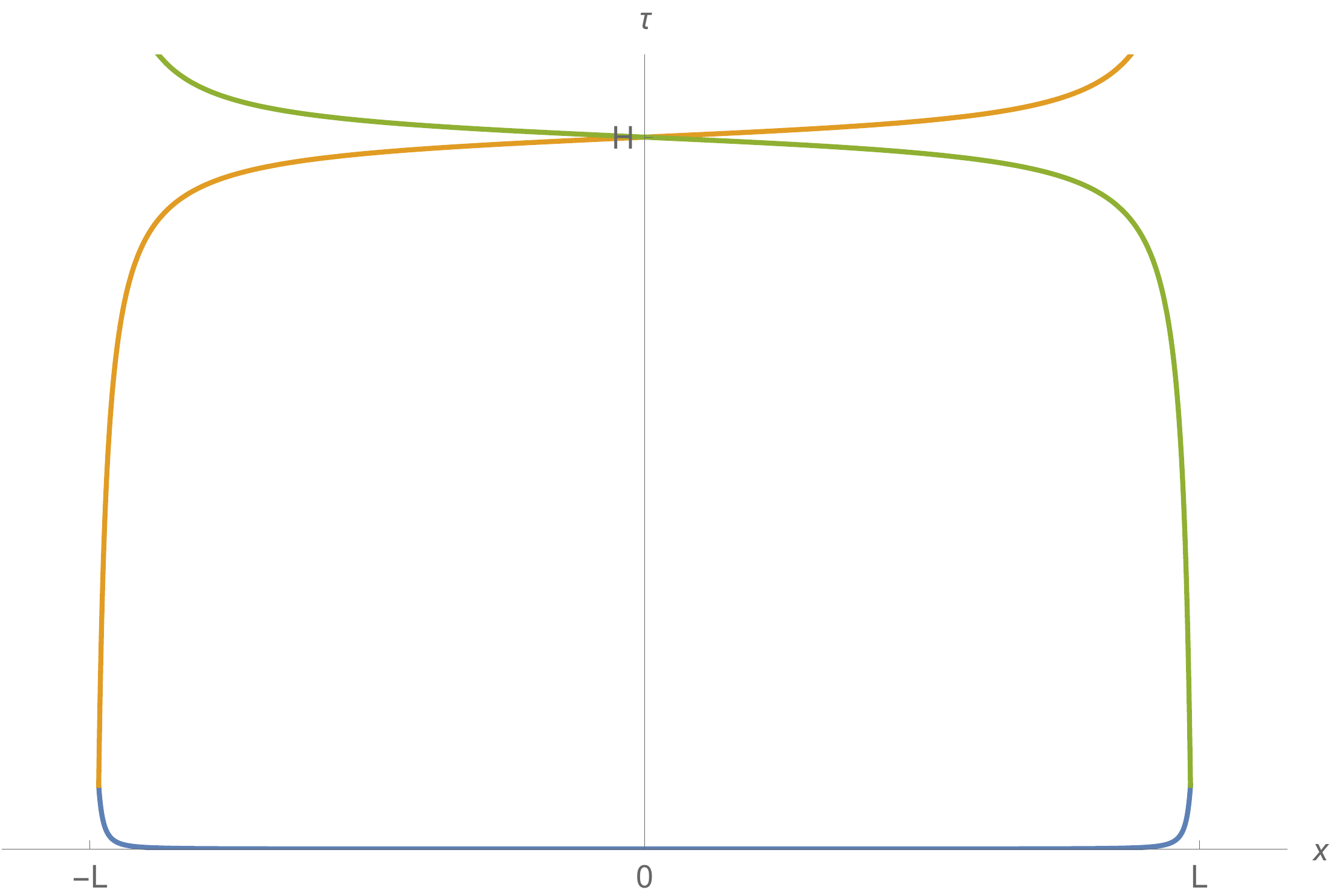}
    \caption{The spacetime boundary of the optimized path integral for the SSD deformed background on the interval $x\in [ -L, L ]$. If one treats the boundary conditions carefully, and solves for the boundary cosmological constant $\mu_B $, one finds that the spacelike portions of the boundary $\tau=F_-(x)$ (orange) and $\tau=F_+(x)$ (green) intersect with each other, yielding a finite domain of path integration.}
    \label{fig:SSD_bdy}
\end{figure}

However, if we carefully analyze the boundaries, as discussed in the previous section, we find that after the optimization, they come close to each other and meet in the middle (Fig.\,\ref{fig:SSD_bdy}) making the optimal region of the path integral geometry closed. We can then evaluate the complexity action in this new region, including the boundary terms parameterized by $\mu_B$. A natural expectation from the SSD deformation would be that the total answer now is independent of the boundary condition $\mu_B$ (see appendix \ref{SSDBCFT}). The computation is straightforward and we can see that almost all terms with $\mu_B$ dependence (that only enters via $F_\pm(x)$) cancel. However we are still left with a contribution that not only depends on $\mu_B$ but also modifies the leading divergent piece in the path integral complexity. One way to understand this result is the fact that the point where two boundaries meet at $x=0$ is equal to
\be
H=\frac{\pi\epsilon+L\alpha \sin\left(\frac{\pi\epsilon}{L}\right)}{2\pi \sin^2\left(\frac{\pi \epsilon}{2L}\right)}\simeq \frac{2L^2(1+\alpha)}{\pi^2\epsilon}+O((1-\alpha)\epsilon).
\ee
This is divergent as $\epsilon\to 0$ in the same way as our leading contribution and indeed our integrals affect it.

We expect that to fully cancel the dependence on $\mu_B$ one should carefully treat the point (cusp) where two boundaries intersect and regulate it appropriately to make it smooth. Perhaps an addition of a Hayward-type term \cite{Hayward:1993my} could cure the problem, but a thorough analysis is beyond the scope of this work and we plan to return to this issue in the future. We hope that the above example illustrates the importance of the boundary contributions to the path integral complexity, and that they deserve further study.

%%%%%%%%%%%%%%%%%%%%%%%%
%%%%%%%%%%%%%%%%%%%%%%%%
\section{Conclusions}\label{Concl}
%%%%%%%%%%%%%%%%%%%%%%%%
%%%%%%%%%%%%%%%%%%%%%%%%

In this work, we developed a framework of path integral optimization and path integral complexity in a class of inhomogeneous 2d CFT --- theories obtained by placing CFTs on curved backgrounds. The inhomogeneous metric is specified by one position dependent function $f(x)$ and specific examples of inhomogeneous CFTs studied in the literature correspond to particular choices of $f(x)$ (e.g. M\"obius, SSD, Rainbow deformations).

In particular, by utilizing a uniformizing coordinate transformation, we found various classes of inhomogeneous backgrounds that solve the Liouville equations with background curvature and optimize the path integral complexity. Their novel features are the explicit position dependence (inhomogeneity) and a position dependent cut-off nedded to fulfill the matching with the physical cut-off of the CFT. 

Moreover, we show how each of the optimized 2d curved metrics can be explicitly embedded into three dimensional $AdS$ spacetimes and, for holographic CFTs with large central charge, can be thought of as a particular slice of the holographic bulk geometry. Interestingly, three dimensional Einstein's equations for metrics with our slices become our condition for minimal path integral complexity i.e. the Liouville equation with non-trivial background curvature. Our findings then provide further support for the idea that holographic bulk geometries can be thought of as a collection of tensor networks of path integrations \cite{Takayanagi:2018pml}.

After presenting several explicit solutions (the vacuum, the primary state, and the thermofield double solutions), we computed their path integral complexity for general $f(x)$ as well as for a few specific examples by directly evaluating the on-shell Liouville action. We find that in the vacuum and primary solution cases, the leading divergences of the complexity remain the same (a reflection of the ``hyperbolicity of the networks"). However, subleading $\mathcal{O}(1)$ terms can be modified by the curvature. In the case of the thermofield double, there is an interplay between the curvature scale of the background and the energy scale introduced by the temperature. This can be seen in the constant curvature example, where the curvature directly changes the effective temperature, as witnessed by the path integral complexity. 

There are several possible generalizations and extensions of this work. The obvious one is a more thorough analysis and classification of different examples of $f(x)$ from the perspective of the curvature as well as path integral geometry and complexity. More general background metrics, including random ones, are also of particular and timely interest in the context of $T\bar{T}$ deformations \cite{Cardy:2018sdv}. Understanding the path integral geometry of these interesting deformations, beyond perturbation theory, is a very important open problem. 

However, the most interesting consequences could be seen by considering time evolving systems. Our work suggests that this could be accomplished e.g. by explicitly adding time dependence to the inhomogeneous metrics to produce a quantum quench geometry. This could provide insight into the connection between complexity and chaos/thermalization \cite{2020JHEP...01..134B,2019arXiv191204297B}, which may have implications on the physics of black holes in a dual gravitational theory. Floquet systems are another source of out of equilibrium physics that may be worth studying with path integral optimization. Some recent work has been done on CFTs with periodic driving using inhomogeneous Hamiltonians \cite{2018arXiv180500031W, 2019arXiv190805289F}.

Last but not least, understanding the mechanism behind AdS/CFT --- how exactly CFT states encode holographic information about $AdS$ metrics --- is the most important task. The new insights from path integral optimization and spacetimes as slices of quantum circuits are still among the first steps in this direction, but developing them further, especially in dynamical setups, will keep us occupied for some time to come.

%%%%%%%%%%%%%%%%%%%%%%%%
%%%%%%%%%%%%%%%%%%%%%%%%
\section*{Acknowledgments}
%%%%%%%%%%%%%%%%%%%%%%%%
%%%%%%%%%%%%%%%%%%%%%%%%
We would like to thank Shouvik Datta, Michal Heller, Jonah Kudler-Flam, Milosz Panfil, Onkar Parrikar, Shinsei Ryu, Joan Simon for discussion and comments on the draft and especially Tadashi Takayanagi and Kento Watanabe for many crucial comments and suggestions during this project. PC would like to thank The Institute for Advanced Study and AEI Golm where parts of this work were done and presented. PC is supported by NAWA ``Polish Returns 2019" grant.

\begin{appendix}
%%%%%%%%%%%%%%%%%%%%%%%%%%%%%%%%%%%%%%%%%%%%%%%%%%%%%
%%%%%%%%%%%%%%%%%%%%%%%%%%%%%%%%%%%%%%%%%%%%%%%%%%%%%
\section{M\"obius/SSD deformed Lagrangians}\label{appendix:LFS}
%%%%%%%%%%%%%%%%%%%%%%%%%%%%%%%%%%%%%%%%%%%%%%%%%%%%%
%%%%%%%%%%%%%%%%%%%%%%%%%%%%%%%%%%%%%%%%%%%%%%%%%%%%%
In this appendix we give a brief derivation of the inhomogeneous ``deformation" in the Lagrangian approach. We start with the analysis of the free boson done in \cite{Tada:2014wta}, where it was shown that a M\"obius deformation leads to the Lagrangian 
\be
\mathcal{L}_\lambda=\frac{1}{2}\int^L_0 dx \left[F(x)\partial_t\varphi^2-G(x)\partial_x\varphi^2\right],
\ee
where
\be
F(x)=N\sum_{k\in\mathbb{Z}}r^{|k|}e^{\frac{2\pi i k x}{L}},\qquad G(x)=1-\lambda \cos\frac{2\pi x}{L},
\ee
and it was found that
\be
r=\frac{1-\sqrt{1-\lambda^2}}{\lambda},\qquad N=\frac{1}{\sqrt{1-\lambda^2}}.
\ee
We can do this sum explicitly and confirm that
\be
F(x)=\frac{1}{1-\lambda\cos\frac{2\pi x}{L}}=\frac{1}{f(x)},\qquad G(x)=f(x),
\ee
where in the conventions of the main text \eqref{MdF}, $\lambda=\tanh(2\gamma)$.\\
This then yields the free boson action in a curved background
\be
S_\lambda=-\frac{1}{2}\int dt\int^L_0 dx\sqrt{g}\left(g^{\mu\nu}\partial_\mu\varphi\partial_\nu\varphi\right)
\ee
where
\be
ds^2=-f(x)^2dt^2+dx^2.
\ee
Let us now elaborate on this geometric deformation in the Euclidean signature. Starting from the free scalar in complex coordinates
\be
S_0=\frac{1}{2}\int d^2z\sqrt{g}g^{\mu\nu}\partial_\mu\varphi\partial_\mu\varphi=\int d^2z\,\partial\varphi\bar{\partial}\varphi,\label{UDFS}
\ee
we have the energy-momentum tensor 
\be
T_{\mu\nu}=-\frac{2}{\sqrt{|g|}}\frac{\delta S_0}{\delta g^{\mu\nu}}=-\left[\partial_\mu\varphi\partial_\nu\varphi-\frac{1}{2}g_{\mu\nu}g^{\alpha\beta}\partial_\alpha\varphi\partial_\beta\varphi\right]=-\left[\partial_\mu\varphi\partial_\nu\varphi-2g_{\mu\nu}\partial\varphi\bar{\partial}\varphi\right]
\ee
such that $T_{z\bar{z}}=T_{\bar{z}z}=0$ and
\be
T_{zz}=-\partial \varphi^2=J(z)J(z),\qquad T_{\bar{z}\bar{z}}=-\bar{\partial} \varphi^2=\bar{J}(\bar{z})\bar{J}(\bar{z}),
\ee
where we defined currents $J(z)=i\partial \varphi$ and $\bar{J}(\bar{z})=i\bar{\partial}\varphi$.

Next, we put this theory on a general curved background
\be
ds^2=f(x)^2dt^2+dx^2=\frac{1}{4}(f^2-1)(dz^2+d\bar{z}^2)+\frac{1}{2}(1+f^2)dzd\bar{z}, \label{MetEU}
\ee
where we have introduced complex coordinates $z=t+ix$. Then the action on this metric can be written as
\be
S_\lambda=S_0+\int d^2z\frac{f-1}{4f}\left[J^2_++f\,J^2_-\right]
\ee
where $L_0$ is the undeformed free scalar part \eqref{UDFS} and we introduced $J_\pm=J\pm \bar{J}$.

As in the case of M\"obius transformations, we can write $f$ as $f=1-4\lambda g$ and then
\be
S_\lambda=S_0-\lambda \int d^2z g\left[J^2_-+\frac{1}{1-4\lambda g}J^2_+\right]
\ee
 Finally we can rewrite it as
\be
S_\lambda=S_0-\lambda \int d^2z g\,J^2_-- \sum^\infty_{k=1}4^{k-1}\lambda^k\int d^2zg^k J^2_+
\ee
From this perspective, we can see that introducing the curved background \eqref{MetEU} can be seen as deforming the original free scalar action by current operators $J^2_\pm$ with position dependent couplings. It will be interesting to develop this observation further and possibly link it with complexity discussions in \cite{Bhattacharyya:2018wym}.

%%%%%%%%%%%%%%%%%%%%%%%%
%%%%%%%%%%%%%%%%%%%%%%%%
\section{Maps to Poincare coordinates}\label{MapsP}
%%%%%%%%%%%%%%%%%%%%%%%%
%%%%%%%%%%%%%%%%%%%%%%%%
Here we explicitly write down the map between $AdS_3$ in Poincare coordinates
\begin{equation}
ds^2=l^2\frac{d\tau^2+dx^2+d\eta^2}{\eta^2},   \label{} 
\end{equation}
and our ansatz metric
\begin{equation}
 ds^2=l^2(d\rho^2+\cosh^2(\rho) e^{2\phi(z,\bar{z})}dzd\bar{z}),
\end{equation}
with the solution of the Liouville equation
\begin{equation}
e^{2\phi(z,\bar{z})}=\frac{4A'(z)B'(\bar{z})}{(A(z)+B(\bar{z}))^2}=\frac{4A'(z)\tilde{B}'(\bar{z})}{(1-A(z)\tilde{B}(\bar{z}))^2},\label{MetPar}
\end{equation}
and coordinates given by
\begin{equation}
\sinh\rho=\frac{x}{\eta},\quad A(z)+B(\bar{z})=2\sqrt{x^2+\eta^2},\qquad A(z)-B(\bar{z})=2i\tau.    
\end{equation}
It can also be written as
\begin{equation}
\eta=\frac{A(z)+B(\bar{z})}{2}\frac{1}{\cosh\rho},\qquad x=\frac{A(z)+B(\bar{z})}{2}\tanh\rho,\qquad \tau= \frac{A(z)-B(\bar{z})}{2i}.\label{TheMap}   
\end{equation}
The parametrization of the general solution of the Liouville equation can be obtained by starting from the ``seed" solution metric
\be
ds^2=e^{2\phi(z,\bar{z})}dzd\bar{z}=\frac{4dzd\bar{z}}{(z+\bar{z})^2}
\ee
and sending $z\to A(z)$ and $\bar{z}\to B(\bar{z})$. The relation to our solution \eqref{GenSolF}, or the second equality in \eqref{MetPar}, is simply $B=-1/\tilde{B}$ in the maps above.

We can now easily check that if we take $z=t+iy$ as in \eqref{CoordChange}, the map \eqref{TheMap} takes us from Poincare to
\begin{equation}
 ds^2=l^2(d\rho^2+\cosh^2(\rho) e^{2\phi(t,x)}(f^2(x)dt^2+dx^2)).
\end{equation}
with $\phi(t,x)$ given by
\be
e^{\phi(t,x)}=\frac{1}{f^2(x)}e^{2\phi(z(t,x),\bar{z}(t,x))},
\ee
which solves the Liouville equation \eqref{EOM} with background metric 
\be
ds^2=f^2(x)dt^2+dx^2.
\ee
Analogous maps can be written for the disc by setting $z=re^{iy}$.

%%%%%%%%%%%%%%%%%%%%%%%%
%%%%%%%%%%%%%%%%%%%%%%%%
\section{BCFT with SSD }\label{SSDBCFT}
%%%%%%%%%%%%%%%%%%%%%%%%
%%%%%%%%%%%%%%%%%%%%%%%%
In this section we present some details of the evaluation of the complexity action for the SSD deformation on a semi-infinite strip $x\in[-L,L]$ with new optimized boundary region.\\
With the new boundaries, we start by evaluating the bulk action (supressing the $x$ dependence)
\bea
S_L[\phi,\hat{g}]&=&\frac{c}{24\pi}\int^0_{-L}dx\int^{F_-}_{\frac{\epsilon}{f}}\left[\frac{2}{\tau^2f}+\frac{f'^2}{f}+2f''\log(\tau f)\right]d\tau\nn\\
&+&\frac{c}{24\pi}\int_0^Ldx\int^{F_+}_{\frac{\epsilon}{f}}\left[\frac{2}{\tau^2f}+\frac{f'^2}{f}+2f''\log(\tau f)\right]d\tau.
\eea
In the formulas we have $F_\pm(x)$ given by \eqref{Fbcft} with \eqref{fSSDs} and \eqref{YSSD}. We will first simplify the general expressions and insert explicit functions at the end.

By first doing $\tau$ integrals we can write the bulk action as a sum of two integrals
\bea
I_1&=&\frac{c}{24\pi}\left[2\left(\frac{L}{\epsilon}-\int^0_{-L}\frac{dx}{fF_-}\right)+\int^0_{-L}dx\frac{f'^2}{f}\left(F_--\frac{\epsilon}{f}\right)\right.\nn\\
&+&\left.2\int^0_{-L}dxf''F_-(\log(fF_-)-1)-2\epsilon(\log(\epsilon)-1)\int^0_{-L}dx\frac{f''}{f}\right],
\eea
and
\bea
I_2&=&\frac{c}{24\pi}\left[2\left(\frac{L}{\epsilon}-\int^L_{0}\frac{dx}{fF_+}\right)+\int^L_{0}dx\frac{f'^2}{f}\left(F_+-\frac{\epsilon}{f}\right)\right.\nn\\
&+&\left.2\int^L_{0}dxf''F_+(\log(fF_+)-1)-2\epsilon(\log(\epsilon)-1)\int^L_{0}dx\frac{f''}{f}\right].
\eea
Next we can simplify the expression in the above formulas on our solution using \eqref{FFP}
\be
2\int^0_{-L}dxf''F_-(\log(fF_-)-1)=2f'(F_-(\log(fF_-)-1))|^0_{-L+\epsilon}-2\int^0_{-L}dxf'(F_-(\log(fF_-)-1))'
\ee
and then
\be
f'(F_-(\log(fF_-)-1))'=\frac{f'^2F_-}{f}+\alpha_-\frac{f'}{f}\log(fF_-),
\ee
and analogously for $I_2$. This way we can write
\bea
S_L[\phi,\hat{g}]&=&\frac{c}{12\pi}\frac{2L}{\epsilon}-\frac{c}{24\pi}\left[\int^0_{-L}dx\frac{f'^2}{f}F_-+\int^L_{0}dx\frac{f'^2}{f}F_+\right.\nn\\
&+&\left.\epsilon\int^L_{-L}dx\frac{f'^2}{f^2}+2\epsilon(\log(\epsilon)-1)\int^L_{-L}dx\frac{f''}{f}\right]\nn\\
&-&\frac{c}{12\pi}\left[\int^0_{-L}dx\left(\alpha_-\frac{f'}{f}\log(fF_-)+\frac{1}{fF_-}\right)+\int^L_{0}dx\left(\alpha_+\frac{f'}{f}\log(fF_+)+\frac{1}{fF_+}\right)\right]\nn\\
&+&\frac{c}{12\pi}f'F_-(\log(fF_-)-1)|^0_{-L+\epsilon}+\frac{c}{12\pi}f'F_+(\log(fF_+)-1)|^{L-\epsilon}_{0},\nn\\
\eea
Note that in the above expression we leave the $O(\epsilon)$ coefficients since the integrals will have to be regulated at $-L+\epsilon$ and $L-\epsilon$ and we only expand at the end.

In the complexity action we also subtract the volume part which is now equal to
\bea
S_L[0,\hat{g}]&=&\frac{c}{24\pi}\int^0_{-L}dxfF_-+\frac{c}{24\pi}\int_0^LdxfF_++O(\epsilon)
\eea
Next we consider the boundary contributions to complexity action. The induced metric for general $\tau=F(x)$ boundaries is given by 
\be
ds^2=\left(1+f(x)^2 F'(x)^2\right)dx^2,\qquad \sqrt{\hat{h}}=\sqrt{1+f(x)^2 F'(x)^2},
\ee
and we can just use this result with $F(x)=\epsilon/f(x)$ or $\tau=F_\pm(x)$ in our computations. This way
\be
\sqrt{\hat{h}}K_{\hat{g}}=-\frac{f'(x) F'_\pm(x) \left(2+f(x)^2 F'_\pm(x)^2\right)+f(x) F_\pm''(x)}{\left(1+f(x)^2 F_\pm'(x)^2\right)},
\ee
so
\be
\sqrt{\hat{h}}K_{\hat{g}}=\epsilon \frac{\frac{f''}{f}+\epsilon^2\frac{f'^4}{f^4}}{1+\epsilon^2\frac{f'^2}{f^2}}
\ee
on $\tau=\epsilon/f(x)$ and
\be
\sqrt{\hat{h}}K_{\hat{g}}=-\alpha_\pm\frac{f'}{f},
\ee
on the boundaries $\tau=F_\pm(x)$ where again we used \eqref{FFP}.

Finally, on the $\tau=\frac{\epsilon}{f(x)}$ boundary we have
\be
\phi=-\log(\epsilon),
\ee
and on $\tau=F_\pm(x)$ we have
\be
\phi=-\log\left(fF_\pm\right).
\ee
This allows us to evaluate all three contributions from the boundaries
\bea
B_1&=&\epsilon\log(\epsilon)\frac{c}{12\pi}\int^L_{-L}dx\left[\frac{\frac{f''}{f}+\epsilon^2\frac{f'^4}{f^4}}{1+\epsilon^2\frac{f'^2}{f^2}}\right],\nn\\
B_2&=&\frac{c}{12\pi}\int^0_{-L}dx\left[\alpha_-\frac{f'}{f}\log\left(fF_-\right)+\frac{1}{fF_-}\right],\nn\\
B_3&=&\frac{c}{12\pi}\int^L_{0}dx\left[\alpha_+\frac{f'}{f}\log\left(fF_+\right)+\frac{1}{fF_+}\right].\\
\eea
In general we could also subtract the boundary ``area"
\bea
S^B[0,\hat{h}]&=&\frac{c}{12\pi}\mu_B\int^0_{-L}\sqrt{1+\alpha^2_-}dx+\frac{c}{12\pi}\mu_B\int^L_{0}\sqrt{1+\alpha^2_+}dx,\nn\\
&=&\frac{c}{12\pi}(2L).
\eea

Note that $B_2$ and $B_3$ precisely cancel the third line of the on-shell bulk action that contains $F_\pm$ and the explicit dependence on $\mu_B$. However the first line of the bulk part (as well as the $S[0,\hat{g}]$) contain integrals with $F_\pm$, and one has to supplement the complexity action with some additional term/procedure in order to smooth out this part and connect the cusp in the boundary.

We can evaluate the remaining integrals with explicit SSD functions analytically and in the small $\epsilon$ expansion. The $F_\pm$ independent integrals are
\be
\epsilon\int^{L-\epsilon}_{-L+\epsilon}dx\frac{f'^2}{f^2}=\frac{2 \pi  \epsilon}{L}  \left(2 \cot \left(\frac{\pi  \epsilon }{2 L}\right)-\pi  \left(1-\frac{\epsilon
   }{L}\right)\right)=8-O(\epsilon)
\ee
\bea
2\epsilon(\log(\epsilon)-1)\int^{L-\epsilon}_{-L+\epsilon}dx\frac{f''}{f}&=&\frac{4\pi\epsilon}{L}(\log(\epsilon)-1)\left(\cot\left(\frac{\pi\epsilon}{2L}\right)-\pi\left(1-\frac{\epsilon}{L}\right)\right)\nn\\
&=&-8(1-\log(\epsilon))+O(\epsilon/L),
\eea
\bea
\frac{c}{12\pi}f'F_-(\log(fF_-)-1)|^0_{-L+\epsilon}+\frac{c}{12\pi}f'F_+(\log(fF_+)-1)|^{L-\epsilon}_{0}&=&\frac{c \epsilon  (1-\log (\epsilon )) }{6 L}\cot \left(\frac{\pi  \epsilon }{2 L}\right),\nn\\
&\simeq&\frac{c (1-\log (\epsilon ))}{3 \pi }+O(\epsilon^2),\nn\\
\eea
and the only boundary (of path integral with $\mu_B=0$)
\bea
B_1&=&\frac{c}{6\pi}\log (\epsilon ) \left(\frac{\pi  \epsilon}{L}  \left(2 \cot \left(\frac{\pi  \epsilon }{2 L}\right)-\pi 
   \left(1-\frac{\epsilon }{L}\right)\right)-\arctan\left(\frac{\pi  \epsilon  }{L}\cot
   \left(\frac{\pi  \epsilon }{2 L}\right)\right)\right),\nn\\
   &=&\frac{c}{6\pi} \left(4-\tan ^{-1}(2)\right) \log (\epsilon)+O(\epsilon)
\eea
Finally the remaining integrals with $F_\pm$ and $\alpha(\mu_B)$ dependence  are
\bea
\int^0_{-L+\epsilon}dx\frac{f'^2}{f}F_-=\int^{L-\epsilon}_{0}dx\frac{f'^2}{f}F_+&=&\frac{\pi ^2 \epsilon  \left(1-\frac{\epsilon }{L}\right)}{L \left(1-\cos \left(\frac{\pi  \epsilon}{L}\right)\right)}+\pi  \left(\alpha -\frac{(1+\alpha ) \epsilon }{L}\right) \cot \left(\frac{\pi  \epsilon }{2 L}\right)\nn\\
&+&4 \alpha  \log \left(\sin \left(\frac{\pi  \epsilon }{2 L}\right)\right),\nn\\
&=&\frac{2L (1+\alpha)}{\epsilon }-2 (\alpha +2)+4 \alpha  \log \left(\frac{\pi  \epsilon }{2 L}\right)+O((1-\alpha)\epsilon),\nn\\
\eea
and the subtracted volume
\bea
S[0,\hat{g}]&=&\frac{c}{24\pi}\frac{ \pi  L \epsilon  \left(1-\frac{\epsilon
   }{L}\right)+L^2 \left(\alpha +\frac{(1-\alpha ) \epsilon
   }{L}\right) \sin \left(\frac{\pi  \epsilon }{L}\right)}{\pi \sin ^2\left(\frac{\pi  \epsilon }{2 L}\right)},\nn\\
   &=&\frac{(1+\alpha) c L}{6 \pi ^3 \epsilon }L^2-\frac{\alpha  c L^2}{6 \pi ^3}+O((1-\alpha)\epsilon).
\eea

Note that both of the above integrals not only depend on $\mu_B$ but also affect the leading divergence in complexity. We believe that there should be a more careful regularization procedure for treating the cusp where the two boundaries meet. In particular, this can then render the path integral complexity for the SSD deformed model on a strip to be independent on the boundary condition labeled by $\mu_B$.

\end{appendix}

\bibliography{main.bib}
\bibliographystyle{ieeetr}
\end{document}